\documentclass[a4paper,11pt]{article}
\pdfoutput=1 

\usepackage{jheppub} 

\usepackage{bm}
\usepackage{tikz}
\usepackage{comment}
\usepackage{appendix}
\usepackage{multirow}
\usepackage{amsmath}
\usepackage{slashed}
\usepackage{braket}
\usepackage{amstext,amssymb}
\usepackage{graphicx}
\usepackage{subcaption}
\usepackage{hyperref}\usepackage[normalem]{ulem}
\usepackage{color}
\usepackage{units}
\newcommand{\mathsym}[1]{{}}

\newcommand{\bav}{\begin{array}{cccc}}

\usepackage{wrapfig}
\definecolor{light-gray}{gray}{0.95}
\usepackage{tcolorbox}

\newcommand{\red}[1]{\textcolor{red}{#1}}

\begin{document} 

\vspace{0.2cm}

\begin{center}
{\Large\bf 



CP violating signal at DUNE in presence of nonstandard interactions and 
the role of second oscillation maxima

}
\end{center}

\vspace{0.1cm}

\begin{center}

 {Rajrupa Banerjee$^\dagger$}~\footnote{rajrupab@iitbhilai.ac.in}, {Jogesh Rout$^\ddagger$}~\footnote{jogesh.rout1@gmail.com}, {Sudhanwa Patra$^\dagger$}~\footnote{sudhanwa@iitbhilai.ac.in}, {Poonam Mehta$^\ast$}~\footnote{pm@jnu.ac.in}
\\
\vspace{0.2cm}
{\it $^\dagger$ Department of Physics, Indian Institute of Technology Bhilai, Durg 491002, India\\
\it $^\dagger$Department of Physics, Shree Ram College, Rampur, Subarnapur, Odisha 767045, India \\
\it $^\ddagger$Institute of Physics, Bhubaneswar, Sachivalaya Marg, Bhubaneswar 751005, India\\
\it $^\star$School of Physical Sciences, Jawaharlal Nehru University, New Delhi 110067, India
}
\end{center}
\vspace{0.05cm}

\begin{center}
\Large{Abstract}
\end{center}
\begin{abstract} 

Neutrino oscillation among the three active neutrino flavors is well established and supported by experiments at diverse length scales and energy scales. It may be noted that five of the neutrino oscillation parameters in the three-flavor paradigm, namely the three mixing angles ($\theta_{12}$, $\theta_{13}$, $\theta_{23}$) and the two mass-squared differences ($\Delta m^{2}_{21}$, $\Delta m^{2}_{31}$) are measured to a reasonable degree of precision. The three unknowns that are expected to be deciphered in the near future are the Dirac CP  phase, $\delta$, the neutrino mass ordering, and the octant of $\theta_{23}$. The next generation of long baseline experiments, such as the Deep Underground Neutrino Experiment (DUNE), aims to resolve these unanswered questions.  
In the present work, by considering DUNE as an example, we assess the ability of long baseline experiments to extricate the intrinsic contribution from observables related to CP violation in scenarios with Standard Interaction (SI) and beyond. Additionally, we analyze the role of the second oscillation maximum in addressing the above mentioned questions. 
By carrying out event level and statistical analyses, we assess the potential of DUNE to probe CP violation effects both within and beyond the standard paradigm.
\end{abstract}

\def\thefootnote{\arabic{footnote}}
\setcounter{footnote}{0}

\newpage

\section{Introduction}
The idea of neutrino oscillations was proposed by Pontecorvo, Maki, Nakagawa, and Sakata~\cite{
Pontecorvo:1957qd,
Pontecorvo:1957cp,
Pontecorvo:1967fh,
Gribov:1968kq,
Maki:1962mu}.
Neutrino oscillations imply that neutrinos possess tiny masses and undergo flavor transitions during propagation. Since neutrinos are assumed to be massless within the Standard Model (SM), the discovery of neutrino oscillations provides us with an unprecedented opportunity to probe physics beyond the SM~\cite{nobel2015}. Much of the data from various neutrino oscillation experiments can be explained by invoking the standard paradigm of three neutrino mixing (see~\cite{
deSalas:2020pgw,
Capozzi:2025wyn,
Esteban:2024eli} for recent global analyses of oscillation data). There are also some anomalous results that cannot be reconciled within the standard three neutrino paradigm. This calls for other possibilities beyond the minimal framework needed to explain neutrino mass and mixing. Among various possible explanations, the idea of non-standard interactions (NSI) of neutrinos (see~\cite{Farzan:2017xzy} for a review) and additional sterile neutrino states (see~\cite{Dasgupta:2021ies} for a review) have been widely explored in the literature. Other scenarios, such as violation of Lorentz invariance and CPT symmetry, have also been invoked~\cite{DUNE:2020fgq}. 

In the standard three neutrino paradigm, the neutrino oscillation framework has nine flavor parameters (three masses, three mixing angles, and three phases) - five of these (three angles and two mass-squared differences) have been measured to a fair degree of precision, as is evident from the global analyses of data~\cite{
deSalas:2020pgw,
Capozzi:2025wyn,
Esteban:2024eli}.
Neutrino oscillations are insensitive to the two Majorana phases and the absolute scale of the neutrino mass. 
The three angles ($\theta_{12},\theta_{13}, \theta_{23}$) and the Dirac CP
phase ($\delta$) appear in the $3 \times 3$ leptonic mixing matrix, commonly referred to as the Pontecorvo-Maki-Nakagawa-Sakata (PMNS) matrix~\cite{ParticleDataGroup:2024cfk}. In vacuum, $\delta$ induces CP and T violation
effects, and we shall refer to these as {{intrinsic or real}} CP effects. Since matter is CP asymmetric, additional CP violating effects may be induced during propagation in matter, making it difficult to infer the true value of the CP phase that appears in the mixing matrix. These effects are termed {{extrinsic or fake}} as they hinder the determination of the intrinsic CP phase. Determining whether CP violation occurs in the leptonic sector is of fundamental importance. This phenomenon could provide a crucial insight into the mechanisms behind the observed matter-antimatter asymmetry in the universe, potentially through the process of leptogenesis~\cite{Fukugita:1986hr, Davidson:2008bu}~. 

Some of the upcoming long baseline neutrino experiments, such as the DUNE~\cite{
DUNE:2015lol,
DUNE:2020jqi,
DUNE:2020ypp},  T2HK~\cite{
Hyper-Kamiokande:2018ofw},
 T2HKK~\cite{
Hyper-Kamiokande:2016srs}, P2O~\cite{
KM3Net:2016zxf, 
Akindinov:2019flp} etc are promising as these will allow us to measure the three unknowns - the value of the Dirac CP phase $(\delta)$, the sign of $(\Delta m^2_{32})$ or the mass ordering and the octant of $\theta_{23}$. 
The primary objective of DUNE is to probe CP violation in the neutrino sector \cite{DUNE:2015lol,
DUNE:2020jqi,
DUNE:2020ypp}. In its experimental configuration, a high purity wide band muon neutrino beam is produced at Fermilab, which traverses a baseline of 1300 km to reach a liquid argon (LAr) far detector situated on axis at the Sanford Underground Research Facility. DUNE uniquely allows us to study the first oscillation maximum as well as the second oscillation maximum of $\nu_\mu \to \nu_e$ oscillation probability~\cite{Bishai:2013yqo, Bass:2013vcg, Rout:2020emr}. This aids in enhanced sensitivity to CP violation effects. 

NSI of neutrinos was first introduced in the seminal paper by Wolfenstein~\cite{Wolfenstein:1977ue}. 
Several phenomenological studies have been carried out exploring the impact of NSI on the determination of CP violation, mass ordering, and octant of $\theta_{23}$ at various experiments, including the long baseline neutrino experiments~\cite{
Grossman:1995wx,
Fornengo:2001pm,
Blennow:2005qj,
Antusch:2008tz,
Meloni:2009ia,
Gago:2009ij,
Adhikari:2012vc,
Agarwalla:2012wf,
Chatterjee:2014gxa,
Choubey:2014iia,
Agarwalla:2014bsa,
Ohlsson:2013nna,
Choubey:2015xha,
Masud:2015xva, 
deGouvea:2015ndi,
Coloma:2015kiu, 
Masud:2016bvp, 
Masud:2016nuj, 
Ge:2016dlx,
Blennow:2016etl, 
Deepthi:2016erc, 
Fukasawa:2016lew,
Liao:2016orc, 
Agarwalla:2016fkh,
Agarwalla:2016mrc,
Rout:2017udo, 
Masud:2017bcf,
Medhi:2021wxj,
Chatterjee:2021wac,
Brahma:2022xld,
Majhi:2022fed,
Sarker:2023qzp,
Sarker:2024ytu,
Bezboruah:2024yhk,
Medhi:2024rsi}. 
In the present work, we shall restrict ourselves to the NSI only, though there is a body of work dealing with other new physics scenarios such as sterile neutrinos~\cite{Dasgupta:2021ies}. In theoretical frameworks, nine independent non-standard interaction (NSI) parameters, each accompanied by an associated complex phase, are expected to exist in nature. Current experimental data from various neutrino oscillation experiments \cite{Super-Kamiokande:2011dam, MINOS:2013hmj, DUNE:2020fgq} provide constraints on the allowed magnitudes and phases of these NSI parameters \cite{Biggio:2009nt, Biggio:2009kv, Escrihuela:2011cf, Coloma:2023ixt}.

It is known that there are inherent difficulties in elucidating whether  CP is violated in the neutrino sector, even in the case where there are no new interactions. In the presence of NSI~\cite{Masud:2015xva, Masud:2016bvp} or sterile neutrinos~\cite{Parveen:2023ixk, Parveen:2024bcc}, the problem of disentangling is more pronounced (see, for instance,~\cite{Masud:2017bcf, Parveen:2023ixk} for separation of physics scenarios employing different beam tunes at DUNE).
Possible ways have been suggested which allow us to separate the intrinsic and extrinsic sources of CP violation both in the presence of NSI and light sterile neutrinos~\cite{Rout:2017udo, Parveen:2023ixk}. 
In the above backdrop, it is evident that one needs to re-look at these questions, keeping in mind the experimental flexibility achievable, for instance, being able to access the second oscillation maxima in addition to the first. In the present work, we (a) study the impact of NSI on CP violation by examining the role played by the first and second oscillation maxima, (b) address the question of disentangling the intrinsic effects from extrinsic effects in the standard paradigm as well as NSI, and (c) assess the ability of considered long baseline experiments to distinguish between the two scenarios (standard versus NSI).

The structure of this paper is as follows. In Section \ref{sec:2}, we outline the theoretical framework. This includes a review of the electron neutrino appearance probability $P_{\mu e}$ in vacuum, in matter with Standard Interactions (SI) and in matter with NSI. We also mention the  observables to extract the intrinsic contribution to the CP violating phase.
In Section~\ref{sec:3}, we provide experimental details including detector specifications and other  inputs used in our analysis for  DUNE. In Section \ref{sec:4}, we present the expected event rates for the DUNE experiment. In Section~\ref{sec:5}, we describe our results pertaining to  the sensitivity to the CP violating phase and resolution of $\delta$.  Finally, we conclude in Section \ref{sec:6}. 
\section{Theoretical Framework}
\label{sec:2}
NSI of neutrinos can lead to neutrino oscillations even for massless neutrinos~\cite{Wolfenstein:1977ue}. 
Most of the phenomenological studies on neutrino oscillations pertain to the case of vector NSI~\cite{Masud:2015xva, Masud:2016bvp, Masud:2016nuj, Rout:2017udo, Masud:2017bcf, Rout:2018txs,Falkowski:2021bkq, Kling:2025zsb} (see~\cite{Farzan:2017xzy, DUNE:2020fgq} for reviews) 
while in recent times, the case of scalar NSI has also been investigated~\cite{Medhi:2021wxj, Sarker:2023qzp, Sarker:2024ytu, Bezboruah:2024yhk, Medhi:2024rsi}. In the present work, we shall consider vector NSI. Typically, NSI arises from charged-current (CC) or neutral-current (NC) processes.  Due to the stringent constraints on parameters governing CC NSI~\cite{Falkowski:2021bkq,Falkowski:2019kfn, Kopp:2025ffx,Agarwalla:2016fkh,Coloma:2023ixt}, we shall only consider NC NSI~\cite{Chatterjee:2014gxa}. 
The NC NSI Lagrangian is given by
\begin{eqnarray}
    \mathcal{L}^{NC}_{NSI} &=& 2\sqrt{2}\mbox{G}_{F}\sum_{\alpha,\beta,P}\epsilon^{f,P}_{\alpha,\beta}\left(\bar{\nu}_{\alpha}\gamma^{\mu}P\nu_{\beta}\right)\left(\bar{f}\gamma_{\mu}P f\right)\,,
\label{eq:4}
\end{eqnarray}
where $P\in \{P_{R}, P_{L}\}$, $P_{R, L}=\left(1\mp \gamma^{5}\right)/2$, $P_{R}$ and $P_{L}$ are the right and left handed chirality operators, respectively. $\alpha$ and $\beta$ correspond to  neutrino flavors   
and $f\in\{e,u,d\}$ denotes the possible fermion fields.  


In the three flavor framework, the evolution equation for neutrinos in the presence of NSI  is
\begin{eqnarray}
    \hat{H} &=& \left[\frac{1}{2E} U \begin{pmatrix}
0 & 0 & 0  \\
0 & \Delta m_{21}^2 & 0 \\
0 & 0 & \Delta m_{31}^2
\end{pmatrix} U^{\dagger} + V_{CC}
\begin{pmatrix}
1 + \varepsilon_{ee}  & \varepsilon_{e\mu}      & \varepsilon_{e\tau}   \\
\varepsilon_{e\mu}^*  & \varepsilon_{\mu\mu}    & \varepsilon_{\mu\tau} \\
\varepsilon_{e\tau}^* & \varepsilon_{\mu\tau}^* & \varepsilon_{\tau\tau}
\end{pmatrix}\right]
\label{eq:5}~.
\end{eqnarray}
Here, $\Delta m_{ij}^2 = m_i^2-m_j^2$ denotes the mass-squared difference  with  $i,j=1,2,3$ ($i \neq j$), $E$ denotes the neutrino energy, and $V_{CC}$  is the charged current (CC) matter potential defined by
\begin{equation}
V_{CC} = \sqrt{2} G_F N_e \simeq 7.6 \times Y_e \times \frac{\rho}{10^{14}
\mbox{g}/\mbox{cm}^3} \, \mbox{eV}\,,
\label{eq:3}
\end{equation}
where $G_F$ is the Fermi coupling constant,  $\rho$ is the matter density, and 
$Y_e = {N_e}/{(N_p + N_n)}$ is the 
electron fraction. $N_e$ is the electron number density while $N_p$ and $N_n$ are the proton and neutron number densities respectively. For an electrically neutral and isoscalar medium, we have $Y_e \simeq 0.5$. 

The commonly adopted form of the mixing matrix $U$ (referred to as the PMNS mixing matrix) is~\cite{ParticleDataGroup:2024cfk}
\begin{equation*} 
U =\begin{pmatrix} 
 c_{12} c_{13} & s_{12} c_{13} & s_{13}e^{-i\delta_{CP}}\\
 -s_{12} c_{23} -c_{12} s_{13} s_{23}e^{i\delta_{CP}} & c_{12} c_{23} -s_{12} s_{13}s_{23}e^{i\delta_{CP}} & c_{13} s_{23}\\
 s_{13} s_{23}-c_{12} s_{13} c_{23}e^{i\delta_{CP}}   & -c_{12} s_{23} -s_{12} s_{13}c_{23}e^{i\delta_{CP}}  & c_{23} c_{13}  
 \end{pmatrix},
 \end{equation*}
 where, $c_{ij}= \cos{\theta_{ij}}$ and $s_{ij}=\sin{\theta_{ij}}$. 
%
 The new NC interaction terms can affect the neutrino oscillation physics by flavour changing 
or  flavour preserving  interactions. Even though the NSI coupling of the neutrino can be to $e, u, d$, we note that at a phenomenological level,
only the sum (incoherent) of all these individual contributions contributes to the coherent forward scattering of neutrinos on matter.  Normalizing to $N_e$, the effective NSI parameter for neutral Earth matter
is
\begin{eqnarray}
    \epsilon_{\alpha\beta}^{}(x) &=& \epsilon^{e}_{\alpha\beta}+\left(2+\dfrac{N_{n}}{N_e} \right)\epsilon^{u}_{\alpha\beta}+\left(1+2 \dfrac{N_{n}}{N_e}\right)\epsilon^{d}_{\alpha\beta}  = \epsilon^{e}_{\alpha\beta}+ 3 \epsilon^{u}_{\alpha\beta} + 3 \epsilon^{d}_{\alpha\beta}
    \,, 
\label{eq:6}
\end{eqnarray}
where $N_f$ is the density of fermion $f$ in medium crossed by the neutrino and $n$ refers to
neutrons. Also,  these NC type NSI matter effects
are sensitive to the vector sum of NSI couplings.

In Table~\ref{oscparams}, we list the best-fit values of 
 standard neutrino oscillation parameters and their allowed ranges obtained from the global fits~\cite{}. We also list the central values and allowed ranges of NSI parameters.

 \begin{table}[]
\small
{
 \hspace{0cm}
	\begin{tabular}{|c|c|c|c|}
	\hline
    \multicolumn{4}{|c|}{Standard oscillation parameters} \\
        \hline
        \hline
		Parameter   & Best-fit value & 3$\sigma$ range & Relative\\
	   & NO (IO) &  NO (IO) & uncertainity\\
		\hline
		\hline
		$\theta_{12}$ & $33.68^{\circ} (33.68^{\circ})$ & $31.63^\circ\rightarrow 35.95^\circ$ & 2.1\% \\
		$\theta_{13}$ & $8.52^{\circ} (8.58^{\circ})$ & $8.18^\circ\rightarrow 8.87^\circ (8.24^\circ\rightarrow 8.91^\circ)$ & 1.3\%(1.3\%)\\
		$\theta_{23}$  &  $48.5^{\circ}(48.6^\circ)$ & $41.0^\circ\rightarrow 50.5^\circ $ & 3.4\%\\
		$\Delta m_{21}^2  (\rm eV^2)$  & $7.49 \times 10^{-5} (7.49 \times 10^{-5})$ & $6.92\rightarrow 8.05$ & 2.5\%\\
		$\Delta m_{31}^2 (\rm                eV^2)$  & $2.534 \times 10^{-3}(-2.484 \times 10^{-3})$ & $2.451\rightarrow 2.578(-2.547\rightarrow -2.421)$ & 0.8\%(0.8\%)\\
		\hline \hline
         \multicolumn{4}{|c|}{GLOB-OSC with NSI in ES+$\nu$NS} \\
         \hline
        \hline
		NSI Parameter & Central value & 2$\sigma$ range & 3$\sigma$ range \\
		\hline
		\hline
		$\varepsilon_{e\mu}$ & $0.05$ & $-0.12\rightarrow +0.011$ & $-0.18\rightarrow +0.08$ \\
		$\varepsilon_{e\tau}$ & $-0.05$ & $-0.16\rightarrow +0.083$ & $-0.25\rightarrow +0.33$\\
        $\varepsilon_{ee}$ & $-$ & $-0.19\rightarrow +0.20$ & $-0.23\rightarrow +0.25$ \\
		\hline \hline
	\end{tabular}
	\caption{Best-fit and 3$\sigma$ ranges of standard oscillation parameters~\cite{Esteban:2024eli}
    and NSI parameters~\cite{Choubey:2014iia, Coloma:2019mbs, Coloma:2023ixt, Majhi:2022fed}. 
    }. 
	\label{oscparams}
   }
\end{table}
\subsection{A review of $\Delta P_{\mu e}^{CP}$ in presence of NSI}
\label{subsec:2.1}
Using perturbative expansions~\cite{Chatterjee:2021wac, Chatterjee:2014gxa, Meloni:2009ia}, the oscillation probability can be approximated as the sum of three leading-order contributions,
\begin{eqnarray}
    P_{\mu e} & \simeq\, P_0 + P_1 + P_2~,
\label{eq:7}
\end{eqnarray} 
where $P_{0}$ and $P_{1}$ correspond to the standard three flavor oscillation contribution, while $P_{2}$ corresponds to NSI contribution.
Expanding the probability, retaining terms up to third order, one can obtain the expression for the leading order contribution to the probability in the presence of NSI~\cite{Barger:2001yr, Agarwalla:2016fkh}.
\begin{eqnarray}
	\label{eq:8}
	& P_{\rm {0}} &\simeq\,  4 s_{13}^2 s^2_{23}  f^2\,,
    \nonumber
    \\
	\label{eq:9}
	& P_{\rm {1}} &  \simeq\,   8 s_{13} s_{12} c_{12} s_{23} c_{23} \alpha f g \cos\left({\Delta + \delta}\right)\,,
    \nonumber
    \\
	\label{eq:10}
	& P_{\rm {2}} &\!\!  \simeq\,  8 s_{13} s_{23} v |\varepsilon|   
	[a f^2 \cos\left(\delta + \phi\right) + b f g\cos\left(\Delta + \delta + \phi\right)]\,,
\label{eq: component_probability}
\end{eqnarray}
In this expression, the matter potential is characterized by the dimensionless parameter $\left|v\right|=\left|V_{CC}/k_{31}\right|\sim 0.22\left(E/2.5\,\textrm{GeV}\right)$, with $k_{31}\cong \Delta m^{2}_{31}/2E$. The explicit form of the coefficients $a$ and $b$ in Eq.~(\ref{eq: component_probability}) depends on the specific flavor structure of the NSI coupling. The impact of NSIs is determined by the magnitude $\left|\varepsilon\right|$ and the associated phase factor $\phi$ of the complex parameter. In Eq.~(\ref{eq: component_probability}), the parameters $\Delta$, $\alpha$, and $v$ take positive values for the normal mass ordering and negative values for the inverted mass ordering. In this work, we focus exclusively on the normal mass ordering, while a detailed study of the inverted mass ordering is left for future investigation. Furthermore, Eq.~(\ref{eq: component_probability}) applies to neutrinos. For antineutrinos, the signs of the parameters $\delta$ and $v$ are reversed. Given the smallness of the mixing angle $\theta_{13}$ and the parameter $\alpha=\Delta m^{2}_{21}/\Delta m^{2}_{31}=0.03$ \cite{Chatterjee:2021wac} in the case of normal hierarchy, we limit our analysis to a first order approximation in the $\alpha-s_{13}$ expansion. This approach retains the leading contributions and simplifies the analysis, since the higher-order terms are negligible for the small values of $\alpha$ and $\theta_{13}$.
In the present analysis, we have neglected the phase associated with the complex NSI parameter to focus on a simplified framework.
Following~\cite{Agarwalla:2016fkh,Barger:2001yr,Chatterjee:2014gxa}, we introduce the coefficient $f$ and $g$ due to matter effect:
\begin{eqnarray}
	f \equiv \frac{\sin [\left(1-v\right) \Delta]}{1-v}\, ~~\text{and}\qquad  g \equiv \frac{\sin v\Delta}{v}\,.
\label{eq:11}
\end{eqnarray}

\begin{figure}[hbt!]
\centering

\begin{minipage}[b]{0.32\textwidth}
    \centering
    \includegraphics[width=\textwidth]{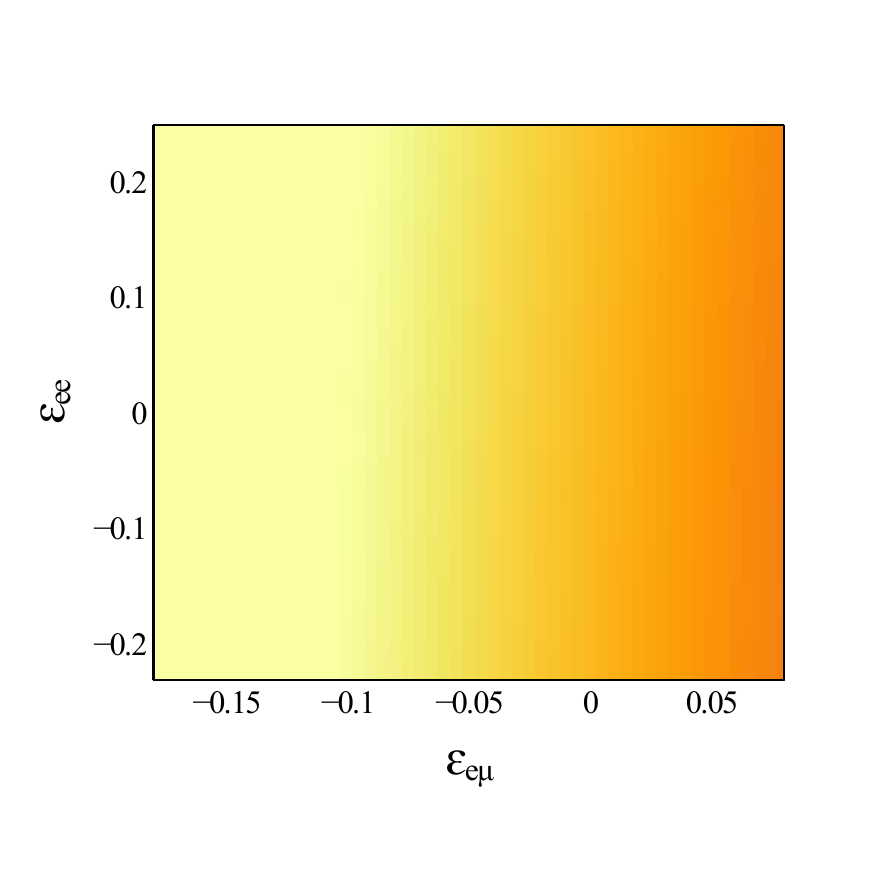}
\end{minipage}
\hfill
\begin{minipage}[b]{0.32\textwidth}
    \centering
    \includegraphics[width=\textwidth]{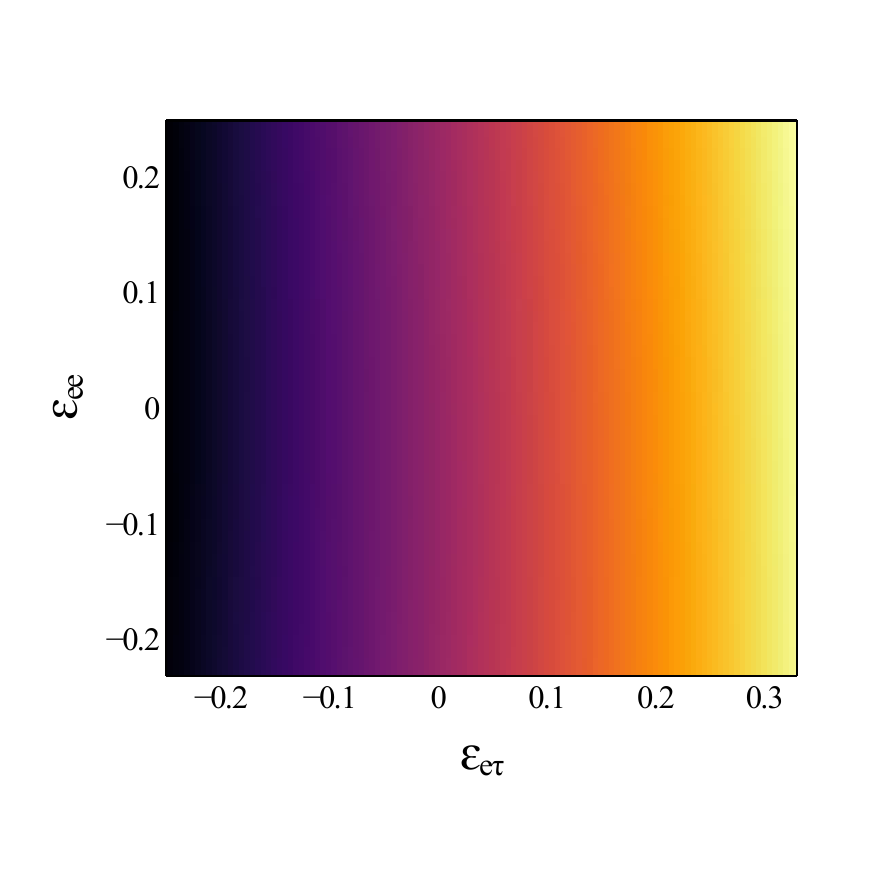}
\end{minipage}
\hfill
\begin{minipage}[b]{0.32\textwidth}
    \centering
    \includegraphics[width=\textwidth]{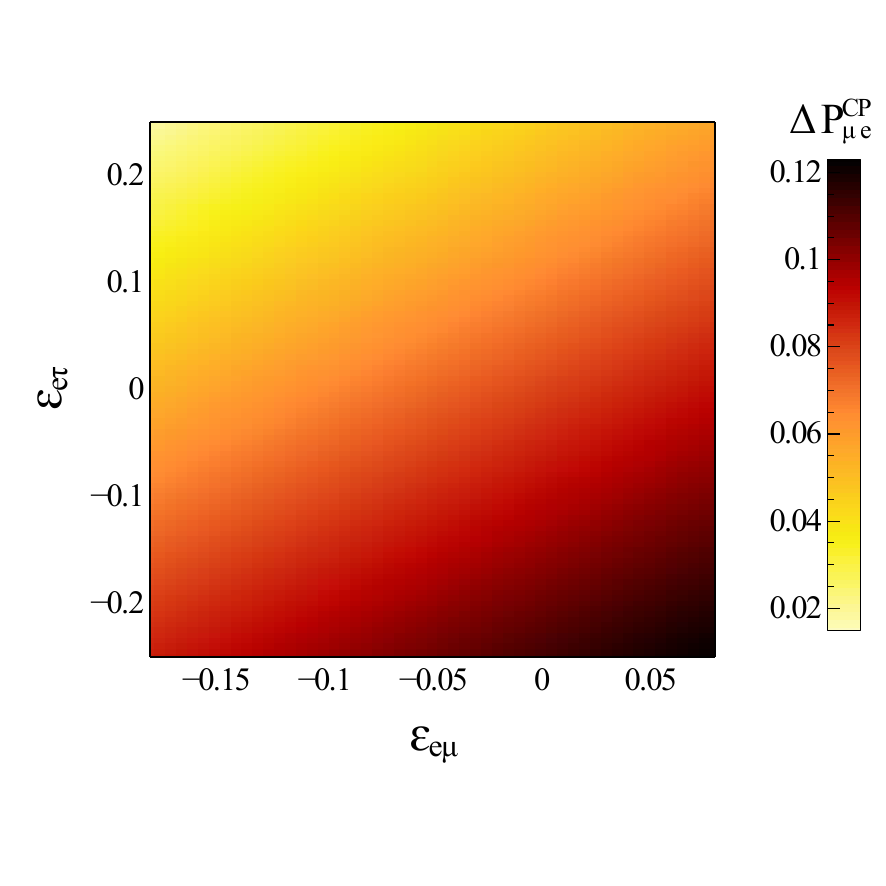}
\end{minipage}
\caption{\footnotesize{{{The oscillogram plot of variation of CP violation paramter $\Delta P^{CP}_{\mu e}$ with different NSI parameter. The left most panel depicts the dependency of $\Delta P^{CP}_{\mu e}$ with $\varepsilon_{ee}$ and $\varepsilon_{e\mu}$, middle panel represents the variation od the CP violation with $\varepsilon_{ee}$ and $\varepsilon_{e\tau}$ and the right panel depicts the same with the variation of $\varepsilon_{e\mu}$ and $\varepsilon_{e\tau}$. }}}} 
\label{fig:3nsiparams}
\end{figure}

The term $P_{0}$ is positive definite, independent of CP phases, and provides the major contribution from the matter effect to the probability. The second term,  $P_{1}$, corresponds to the standard three-flavor interference between the solar and atmospheric oscillation frequencies, and is sensitive to the intrinsic CP violating phase, $\delta$. The third term, $P_{2}$, encapsulates the contribution from non-standard interactions (NSI). This term involves the complex NSI coupling and contributes only in the presence of matter, i.e., when $v\neq 0$. Based on this theoretical framework, we now proceed to determine the relevant NSI parameters and their corresponding values.
It is important to emphasize that all numerical results presented in this work are obtained using the standard neutrino oscillation parameters from the NuFIT v6.0 global fit, as listed in Table~\ref{oscparams}. The non-standard interaction (NSI) parameters are taken within their allowed $3\sigma$ ranges, also summarized in Table~\ref{oscparams}. The choice of the central values for the NSI parameters is motivated by the oscillogram plots shown in Fig.~\ref{fig:3nsiparams} at the first oscillation maximum.
\begin{enumerate}
    \item The leftmost panel displays the variation of $\Delta P^{CP}_{\mu e}$ as a function of $\varepsilon_{ee}$ and $\varepsilon_{e\mu}$. It is evident that $\Delta P^{CP}_{\mu e}$ is largely sensitive to $\varepsilon_{e\mu}$, with a negligible  contribution is arising from $\varepsilon_{ee}$. The enhancement of CP violation is particularly pronounced for positive values of $\varepsilon_{e\mu}$.

    \item The middle panel shows the dependence of \(\Delta P_{\mu e}^{CP}\) on \(\varepsilon_{ee}\) and \(\varepsilon_{e\tau}\). Consistent with the previous observation, the plot confirms that \(\Delta P_{\mu e}^{CP}\) remains nearly unaffected by variations in \(\varepsilon_{ee}\), while displaying a significant dependence on \(\varepsilon_{e\tau}\), especially for negative values of it.

    \item The rightmost panel presents the joint dependence of \(\Delta P_{\mu e}^{CP}\) on both \(\varepsilon_{e\mu}\) and \(\varepsilon_{e\tau}\). From this plot, it becomes clear that the benchmark choice \(\varepsilon_{e\mu} = 0.05\) and \(\varepsilon_{e\tau} = -0.05\) provides a representative scenario for probing significant CP violating effects within the three-flavor neutrino oscillation framework. For the CP violation sensitivity analysis, however, these parameters are systematically varied within their respective \(3\sigma\) bounds, in accordance with the most recent global constraints on NSI couplings as given in Table \ref{oscparams}~\cite{Coloma:2023ixt}.

\end{enumerate}
This analysis reveals that the dominant contributions to $\Delta P_{\mu e}^{CP}$ arise from the NSI parameters $\varepsilon_{e\mu}$ and $\varepsilon_{e\tau}$. Considering only the two NSI parameters $\varepsilon_{e\mu}$ and $\varepsilon_{e\tau}$, the explicit forms of the coefficients $a$ and $b$ in Eq.~(\ref{eq: component_probability}) depend on the flavor structure of the NSI coupling.
For the two most relevant off-diagonal NSI parameters, the expressions are given by \cite{Agarwalla:2016fkh}
\begin{eqnarray}
	\label{eq:13&14}
	a = s^2_{23}, \quad b = c^2_{23} \qquad &{\mathrm {if}}& \qquad \varepsilon = |\varepsilon_{e\mu}|e^{i{\phi_{e\mu}}}\,,\\
	a =  s_{23}c_{23}, \quad b = -s_{23} c_{23} \qquad &{\mathrm {if}}& \qquad \varepsilon = |\varepsilon_{e\tau}|e^{i{\phi_{e\tau}}}\,.
\end{eqnarray}
Thus, the structure of $P_2$ differs slightly between the $\varepsilon_{e\mu}$ and $\varepsilon_{e\tau}$ cases, reflecting the distinct interference patterns these couplings introduce in the oscillation probability.  
The dominant term in Eq.~(\ref{eq: component_probability}) depends on $\Delta m^{2}_{31}$. As a result, the key experimental features of a long-baseline $\nu_\mu \rightarrow \nu_e$ appearance measurement are determined by the baseline $L$ and the neutrino energy $E$, through the ratio $1.27\,\Delta m^{2}_{31}L(\mathrm{km})/E(\mathrm{GeV})$. The condition for the oscillation maxima of $P_{\mu e}$ is obtained from the leading term, which yields  
\begin{eqnarray}
    \frac{L}{E} \sim (2n-1)\times 500~\frac{\mathrm{km}}{\mathrm{GeV}},
\label{oscmax}
\end{eqnarray}
Where $n = 1, 2, \ldots$ correspond to the first, second, and higher oscillation maxima occurring at $L/E \sim 500,\, 1500, \ldots~\mathrm{km/GeV}$, respectively. 
For the DUNE with baseline of $L = 1300$ km, the relation of Eq.~(\ref{oscmax}) places the first and second oscillation maxima at approximately  $E_I \simeq 2.6~\mathrm{GeV}, \quad 
E_{II} \simeq 0.86~\mathrm{GeV}~.$ 
\noindent
 \begin{figure}
 \centering
\includegraphics[width=0.8\textwidth] {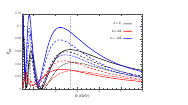}
	\caption{{$\nu_{e}$ appearance probability in vacuum (solid lines), in presence of standard matter effect (dotted lines), and with non-standard interactions (dashed lines) with $\delta = 0, \pi/2, -\pi/2$ considering normal mass ordering. The vertical lines at $2.68\mathrm{GeV}$ and $0.86~\mathrm{GeV}$ referee to the first and the second oscillation maxima. }}
	\label{fig:probability}
\end{figure}
Motivated by these observations, we fix the NSI parameters to $\varepsilon_{e\mu}=0.05$ and $\varepsilon_{e\tau}=-0.05$ and proceed to study the energy dependence of the electron appearance probability, $P_{\mu e}$. The analysis is performed for three different cases: vacuum oscillations (solid curve), standard interactions (dotted curve), and non-standard interactions (dashed curve), with particular emphasis on the first and second oscillation maxima.
The results are shown in Fig.~\ref{fig:probability} for three representative values of the CP phase, $\delta = 0^\circ, +90^\circ,$ and $-90^\circ$. 
The black curve corresponds to $\delta = 0^\circ$, while the red and blue curves illustrate the cases for $\delta = +90^\circ$  and $\delta = -90^\circ$, respectively.
A clear hierarchical behavior is observed in all cases. The probability $P_{\mu e}$ increases successively from vacuum (solid line) to SI (dotted line) and further to NSI (dashed line). This enhancement is more pronounced at the second oscillation maximum than at the first. In particular, for $\delta = -90^\circ$, $P_{\mu e}$ increases from approximately 0.09 in vacuum to about 0.125 in the presence of NSI at the second oscillation maximum. This pronounced rise indicates that the second oscillation maximum offers enhanced sensitivity to CP-violating effects.
Furthermore, the significant modification of $P_{\mu e}$ due to NSI highlights their potential to enhance CP sensitivity beyond that of the standard three-flavor framework.
In the following sections, we examine how this enhancement affects CP sensitivity and discuss its phenomenological implications.
\subsection{$\Delta P^{CP}_{\mu e}$ in NSI}
Using the appearance probability in the presence of NSI given in Eq.~(\ref{eq: component_probability}), we obtain a compact and simplified expression for the CP-violating probability difference when NSI effects are included. The resulting expression is given by
\begin{eqnarray}
     \Delta P^{CP}_{\mu e} &=& \Phi\bigg[\frac{\Phi}{4}\big(f^{2}-f^{\prime^{2}}\big)
	 +v\big(f^{2}+f^{\prime^{2}}\big)\big(\varepsilon_{e\mu}-\xi\big)\cos{\delta} +g(\mbox{x}+\mbox{y})\big(f^{\prime}+f\big)\cos{\Delta}\cos{\delta}
     \nonumber
     \\
     &&
     ~~~~~~~~~~~+g(\mbox{x}-\mbox{y})\big(f^{\prime}-f\big)\sin{\Delta}\sin{\delta}\bigg]
\label{simplifiedCP}
\end{eqnarray}
 where $\Phi=4s_{13}s_{23}, \quad \mathrm{x}=v\left(\varepsilon_{e\mu}+\xi\right), \quad \mathrm{y}=\alpha\sin{2\theta_{12}}\cos{\theta_{23}}, \quad \xi=\varepsilon_{e\mu}\cos{2\theta_{23}}-\varepsilon_{e\tau}\sin{2\theta_{23}}.$ 
For completeness, the similar derivation of $\Delta P^{CP}_{\mu e}$ in both the vacuum and standard interaction (SI) cases is provided in Appendix~\ref{CPvac}.
Eq.~(\ref{simplifiedCP}) reveals that, unlike the SI effect, NSI introduces a $\cos{\delta}$ dependency that is independent of the oscillation maxima. This implies that, at the oscillation maxima, in addition to the sinusoidal dependence on $\delta$ observed in vacuum and SI scenarios, an additional $\cos{\delta}$ dependent term emerges uniquely due to NSI  that restrict $\Delta P^{CP}_{\mu e}$ to approach absolute zero at the CP conserving regions i.e. $\delta=0,\pm \pi$. Since neutrino oscillation experiments primarily focus on the region near the oscillation maximum, one can set $\sin{\Delta} = 1$. Consequently, at the oscillation maximum, Eq.~(\ref{simplifiedCP}) simplifies to the following form,
\begin{eqnarray}
    \Delta P^{CP}_{\mu e} &=& \Phi\left[\frac{\Phi}{4}\left(f^{2}-f^{\prime^{2}}\right)
	 +v\left(f^{2}+f^{\prime^{2}}\right)\left(\varepsilon_{e\mu}-\xi\right)\cos{\delta}+g\left(\mbox{x}-\mbox{y}\right)\left(f^{\prime}-f\right)\sin{\delta}\right].
     \nonumber
     \\
\label{eq:24}
\end{eqnarray} 
However, since the $\cos{\delta}$ term is coupled with both the matter potential and NSI parameters, a reduction in the contribution from the matter effect allows the impact of NSI to be observed more prominently. 
Therefore, in the presence of NSI, the higher oscillation maxima provide an opportunity to reduce the impact of matter effects, thereby improving the extraction of the intrinsic CP phase $\delta$ within the three-flavor neutrino oscillation framework.
\subsection{Extraction of CP violating phase in presence of NSI}
\label{subsec:2.2}
\begin{figure}
	\centering	
\includegraphics[width=.65 \textwidth] {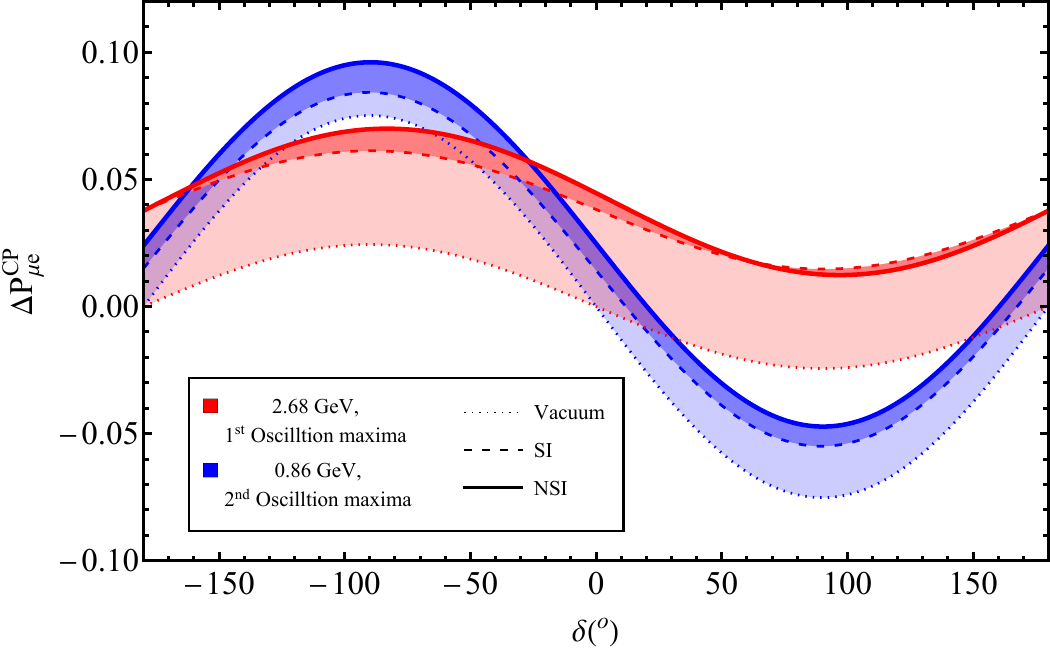}
	\caption{The figure illustrates the variation of the CP violation factor $\Delta P^{CP}_{\mu e}$ as a function of $\delta$ for different oscillation scenarios: vacuum (dotted line), standard matter effects (dashed line), and the presence of NSI (solid line). The red curves correspond to the first oscillation maximum at 2.68 GeV, while the blue curves represent the second oscillation maximum at 0.86 GeV. The shaded region highlights the excess contribution to CP violation arising from matter effects and NSI, beyond the intrinsic CP violation component. The plot demonstrates the amplification of CP violation effects due to matter interactions and NSI, with a stronger influence observed at the second oscillation maximum.}
	\label{fig:3}
\end{figure}

Motivated by the enhanced sensitivity to the CP phase $\delta$ at higher oscillation maxima, we now investigate the possibility of extracting intrinsic CP violation from the observed CP asymmetry, which contains both intrinsic and extrinsic contributions. By separating the intrinsic and extrinsic components and rearranging the terms in the CP asymmetry expression, we obtain
$\Delta P_{\mu e}^{CP} = P_{\mu e} - \bar{P}_{\mu e}$ as shown in Eq.~(\ref{eq:2.25}).
\begin{eqnarray}
	\Delta\textrm{P}^{CP}_{\mu e} 
    &=& \underbrace{4 s_{13}^2 s^2_{23}\left(f^{2}-(f^\prime)^{2}\right)}_{\textrm{Extrinsic}}
    \nonumber
    \\
	&& 
    +~  \underbrace{8 s_{13} s_{23}s_{12} c_{12} c_{23} \alpha g\left[\cos{\delta}\cos{\Delta} \left(f+f^{\prime}\right)+\sin{\Delta}\sin{\delta}\left(f-f^{\prime}\right)\right]}_{\textrm{Extrinsic and Intrinsic CP violation due to matter effect}}
    \nonumber\\
	& & 
    +~\underbrace{8 s_{13} s_{23} v|\varepsilon| \left[a\cos{\delta}\big(f^{2}+f^{\prime^{2}}\big)+b g\left\{\cos{\Delta}\cos{\delta}\left(f+f^{\prime}\right)+\sin{\Delta}\sin{\delta}\left(f^{\prime}-f\right)\right\} \right]}_
    {\textrm{Extrinsic and Intrinsic CP violation due to matter effect and NSI}}~. 
    \nonumber \\
\label{eq:2.25}
\end{eqnarray}
The contributions can be grouped according to their dependence on the CP phase:
\begin{enumerate}
    \item \textit{Term containing $\sin{\Delta}\sin{\delta}$}: These encode a combination of intrinsic and extrinsic CP violation, with contributions originating from both matter effects and NSI. Since the factor $\sin{\Delta}$ peaks at the oscillation maxima, this term remains nonzero at these points and therefore captures the dominant CP violating behavior near the second oscillation maximum.
    \item \textit{Term containing $\cos{\Delta}\cos{\delta}$}: These terms also represent mixed intrinsic-extrinsic CP violation arising from matter and NSI. However, because $\cos\Delta$ vanishes at the oscillation maxima, their contribution is suppressed at these points. As a result, they play no role in the regions of maximal oscillation probability.
    \item \textit{Term containing $\cos{\delta}$}: These contributions are also an entangled effect of both intrinsic and extrinsic CP violation due to matter and NSI effects. Since they do not depend on the oscillation phase $\Delta$, they remain nonzero regardless of the oscillation maxima and are independent of the location of the oscillation peaks.
\end{enumerate}
It is therefore evident that the term proportional to $\sin\Delta,\sin\delta$ survives even in the absence of NSI and constitutes the standard intrinsic source of CP violation. In contrast, the term proportional to $\cos\delta$ arises from NSI through their modification of the matter potential and remains present at all oscillation maxima. However, because matter effects are weaker at the second oscillation maximum, this energy region provides a more favorable setting for observing the impact of the NSI-induced $\cos\delta$ term alongside the $\sin\delta$ term, thereby improving the prospects for extracting the CP phase $\delta$. Having established this theoretical framework, we now proceed to investigate CP violation in the experimental setup of DUNE using the General Long Baseline Experiment Simulator (GLoBES) package~\cite{Huber:2004ka, Huber:2007ji}. The simulations incorporate complete three-flavor neutrino propagation in matter, using the PREM density profile~\cite{Dziewonski:1981xy} along with the oscillation parameter values listed in Table~\ref{oscparams}.
As discussed earlier, the sensitivity to CP violation is significantly enhanced near the second oscillation maximum, as demonstrated by the oscillogram of $\Delta P^{CP}_{\mu e}$ in the $(E,\delta)$ plane.
This feature can be clearly seen in Fig.~\ref{fig:oscideldelp}, which presents the corresponding oscillograms for the vacuum, standard matter, and NSI scenarios.
The orange and blue regions correspond to large CP asymmetries, with magnitudes reaching approximately 0.1, indicating regions of enhanced sensitivity to CP-violating effects.
\begin{enumerate}
    \item \textbf{\textit{Vacuum:}}  
    The leftmost panel of Fig.~\ref{fig:oscideldelp} displays the CP asymmetry in vacuum. As expected from Fig.~\ref{fig:probability}, maximal CP violation occurs at $\delta = \pm 90^{\circ}$, while CP violation vanishes at $\delta = 0^{\circ}$ and $180^{\circ}$. The oscillogram also shows that a large CP asymmetry can occur at higher order oscillation maxima. Although higher maxima provide stronger sensitivity, only the first and second oscillation maxima lie within the accessible energy range of DUNE. In vacuum, the CP asymmetry can reach magnitudes of approximately $0.1$, representing purely intrinsic CP violation, unaffected by matter effects. 
    \item \textbf{\textit{SI:}}  
The middle panel demonstrates how matter effects modify the vacuum pattern. The symmetry observed in vacuum is significantly distorted, particularly around the first oscillation maximum. A region of strong CP asymmetry (visible as a dark orange patch) appears near the first oscillation minimum, a feature entirely absent in vacuum, indicating the role of matter induced (extrinsic) CP violation. The region of large CP asymmetry at the second oscillation maximum also expands in the presence of matter, although a partial symmetry is still retained there. This partial preservation makes the second oscillation maximum particularly valuable for disentangling the intrinsic CP effect from the extrinsic CP effect, maximizing the possibility of extracting $\delta$. 

\item \textbf{\textit{NSI:}}  
\begin{figure}[hbt!]
\centering
\vspace{-1.5cm}
\hspace*{-1.28cm}
\includegraphics[width=0.42\textwidth]{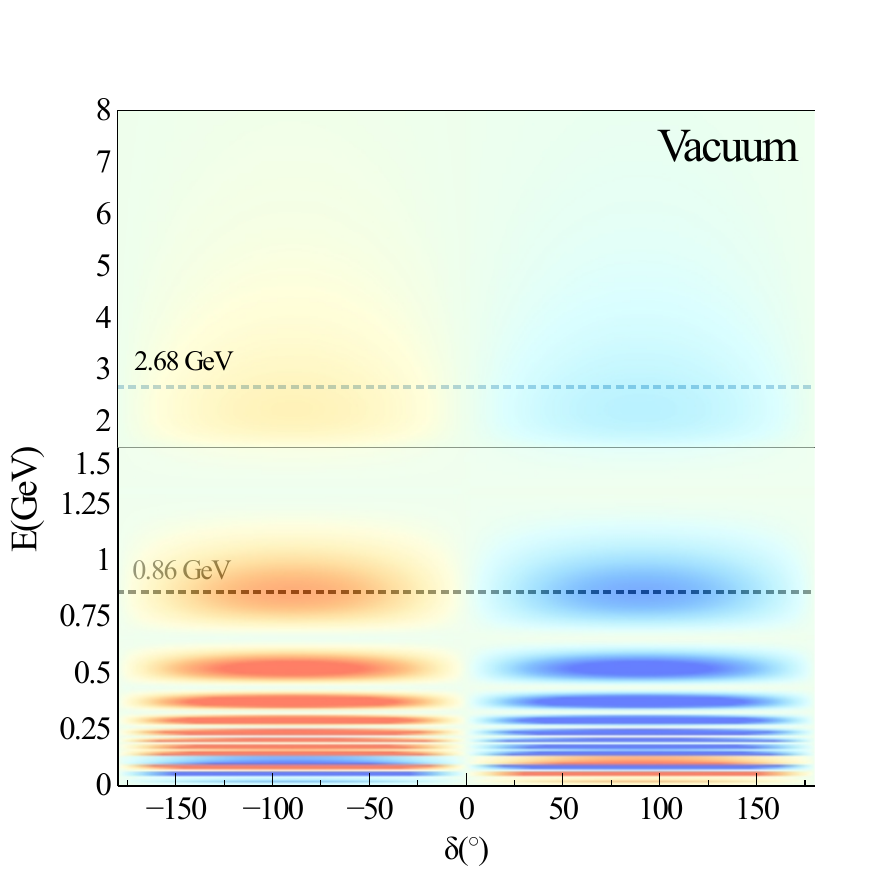}
\hspace{-1.5cm}
\includegraphics[width=0.416\textwidth]{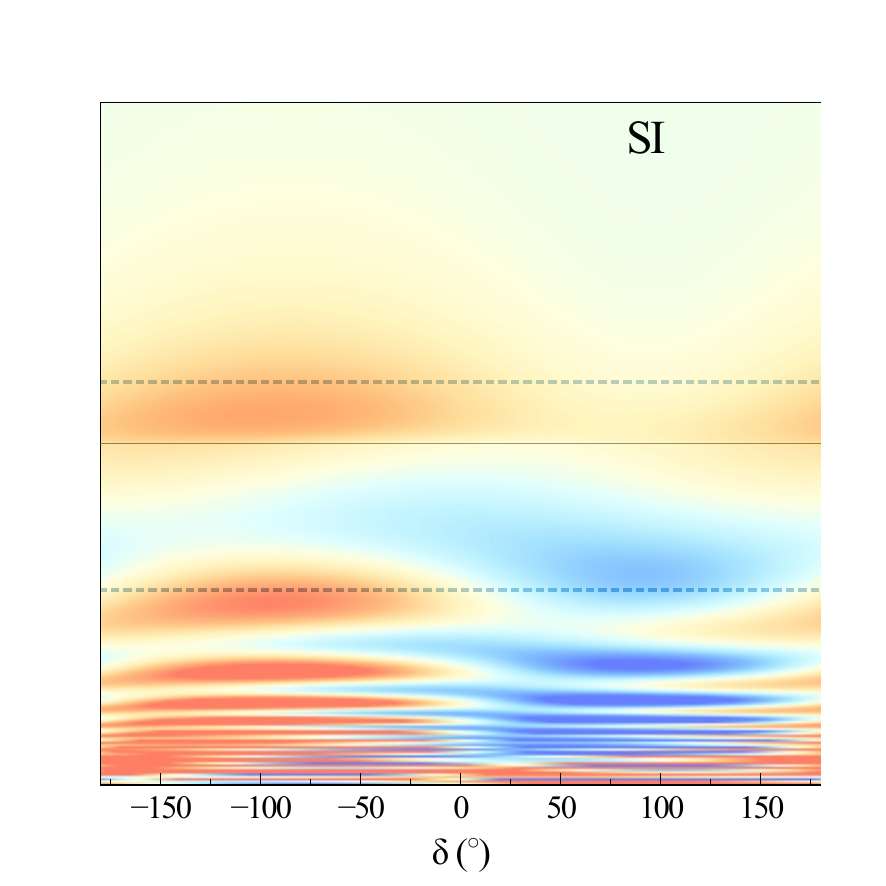}
\hspace{-1.71cm}
\includegraphics[width=0.421\textwidth]{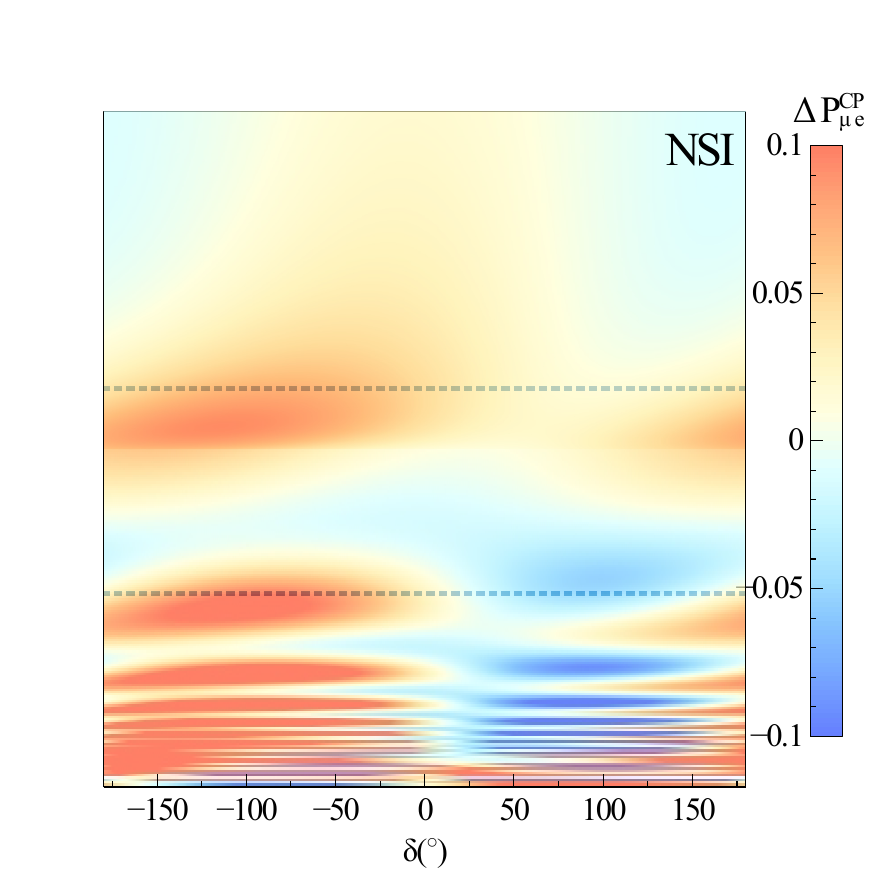}
\caption{\footnotesize{{{The oscillogram plot presented in the figure illustrates the dependence of the CP violation factor $\Delta P^{CP}_{\mu e}$ on the $\delta$. The left panel of the figure represents the scenario in vacuum, where $\Delta P^{CP}_{\mu e}$ exhibits a purely intrinsic CP violation effect, manifesting a symmetric behavior between the maximal and minimal CP violation regions. This symmetry indicates that $\Delta P^{CP}_{\mu e}$ is exclusively governed by the intrinsic CP phase, without external influences. In the presence of standard matter interactions, as shown in the middle panel, the symmetry of $\Delta P^{CP}_{\mu e}$ is significantly distorted. As a result, the extraction of intrinsic CP violation becomes increasingly challenging when transitioning from the vacuum scenario to the matter affected regime. The right panel of the figure further explores the impact of NSI on $\Delta P^{CP}_{\mu e}$. The inclusion of NSI exacerbates the difficulty of isolating the intrinsic CP violation component. Despite this complication, NSI significantly enhances the maximum CP violation values, suggesting that while the extraction of intrinsic CP violation is hindered, the overall magnitude of CP violation is amplified. }}}}  
\label{fig:oscideldelp}
\end{figure}
The rightmost panel shows the oscillogram when NSI are included. NSI causes an even stronger distortion of the symmetry seen in vacuum, especially around the first oscillation maximum, where the symmetry is almost completely broken. The magnitude of CP violation is noticeably enhanced compared to the SI case, reflecting the additional interference introduced by NSI parameters. In contrast, the second oscillation maximum retains partial symmetry and continues to exhibit a pronounced region of significant CP asymmetry. This behavior indicates that the second oscillation maximum not only provides improved sensitivity to large CP violation in the presence of NSI but also offers a more favorable environment for untwining $\delta$ from extrinsic fake CP effect. 
\end{enumerate}
Consequently, the second oscillation maximum emerges as one of the most promising regions for evaluating $\delta$ in long-baseline setups, such as DUNE.
A clearer understanding of these features emerges when examining the two dimensional dependence of $\Delta P^{CP}_{\mu e}$ on the CP phase $\delta$ at the two accessible oscillation maxima. Fig.~\ref{fig:3} illustrates how the CP asymmetry varies with $\delta$ at first and second oscillation maxima, enabling a direct comparison between vacuum, standard matter effects, and NSI contributions. The main observations from the figure are summarized below.
\begin{itemize}
    \item \textbf{Oscillation Maxima:}  
    The red curves correspond to the first oscillation maximum, while the blue curves represent the second. Dotted lines depict vacuum oscillations, dashed lines include standard matter effects, and solid lines incorporate NSI. The light-red(light-blue) shaded bands indicate the shift from intrinsic CP behavior due to matter effects, whereas the dark-red(dark-blue) shades highlight the additional deviation introduced by NSI at first oscillation maximum(second oscillation maximum).
    \item \textbf{Relative Magnitudes:}  
    Across all scenarios, the magnitude of $\Delta P^{CP}_{\mu e}$ is consistently larger at the second oscillation maximum ($E^{II}$) than at the first ($E^{I}$), for both SI and NSI cases.
    \item \textbf{Amplitude Enhancement:}  
    The value of $\Delta P^{CP}_{\mu e}$ at the second maximum is amplified by nearly a factor of three relative to the first in the presence of NSI. This highlights the increased sensitivity of the second maximum to CP violating effects.
    \item \textbf{Peak Behavior:}  
    The largest differences occur near $\delta = \pm 90^\circ$. The maximum value of $\Delta P^{CP}_{\mu e}$ reaches approximately 0.1037 at $\delta = -90^\circ$, observed for both SI and NSI at an energy of  0.86~GeV.

    \item \textbf{CP Conserving Phases:}  
    For CP conserving values ($\delta = 0^\circ$ or $\pm 180^\circ$), $\Delta P^{CP}_{\mu e}$ vanishes in vacuum at both $E^{I}$ and $E^{II}$. In contrast, matter effects and NSI produce a nonzero asymmetry in these regions. In the NSI case, $\Delta P^{CP}_{\mu e}$ reaches 0.049 at $\delta = 0^\circ$ and 0.043 at $\delta = 180^\circ$.

    \item \textbf{Offset Due to Matter:}  
    At the first oscillation maximum, the offset between vacuum and matter induced asymmetry is substantial. This offset becomes noticeably smaller at the second oscillation maximum, making the latter less susceptible to fake CP effects.

    \item \textbf{Symmetry Features:}  
    NSI distorts the symmetry of the CP asymmetry curve around the first oscillation maximum. However, near the second maximum, the NSI induced modification is almost symmetric, effectively adding a nearly constant shift to the matter induced behavior.
\end{itemize}
 The analytical discussion above shows that the second oscillation maximum provides a substantially stronger signal of CP violation compared to the first. This enhancement becomes particularly striking in the presence of NSI.
 \begin{figure}
	\centering	
\vspace{-0.5cm}
\includegraphics[width=.70\textwidth] {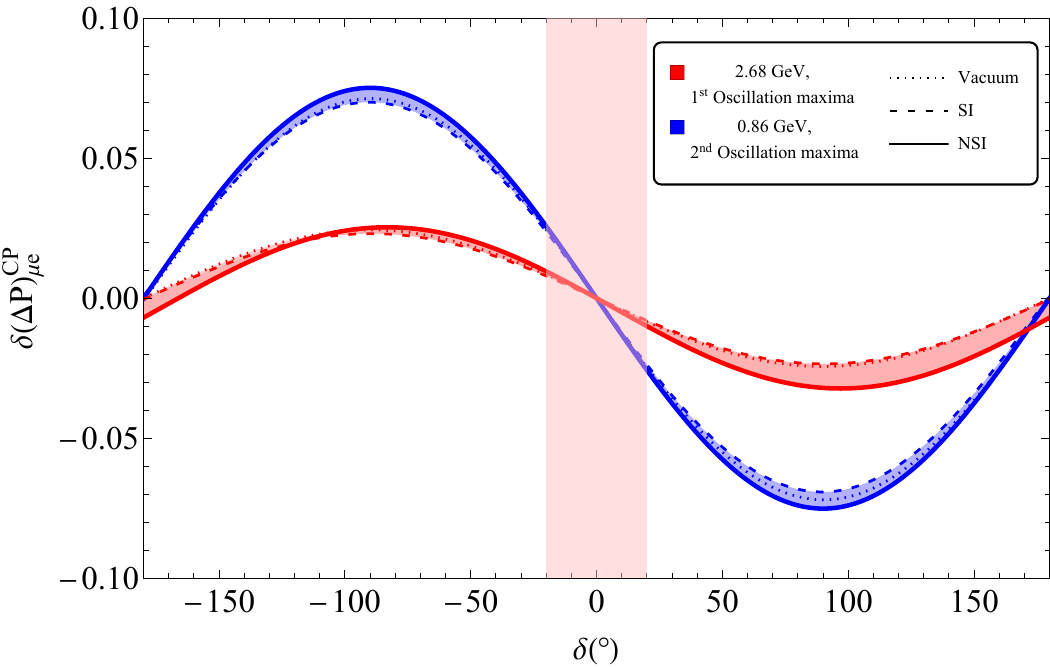}
	\caption{The same plot as Fig.~\ref{fig:3}, now presented with the parameter $\delta\left(\Delta P^{CP}_{\mu e}\right)=\Delta P^{CP}_{\mu e}\left(\delta\right)-\Delta P^{CP}_{\mu e}\left(\delta=0\right)$. This specific parameterization is employed to analyze the contribution of fake CP violation by considering the minimal value of the intrinsic CP phase. The approach allows for a more precise assessment of the impact of matter effects and NSI on CP violation, aiding in the disentanglement of intrinsic and extrinsic contributions~\cite{Majhi:2022fed}.}
	\label{fig:4}
\end{figure} 
\subsection{Importance of higher order oscillation maxima in distinguishing intrinsic CP effects }
\label{subsec:2.3}
Building on the analytical discussion above, it is evident that the second oscillation maximum plays a particularly important role in enhancing the sensitivity to CP violation, especially in the presence of NSI. In this section, we therefore focus on the prospects for extracting the CP phase $\delta$ at the second oscillation maximum.

A useful quantity for distinguishing intrinsic CP violation from extrinsic contributions (arising from matter effects or NSI) is the difference  
\begin{eqnarray}
    \delta\!\left(\Delta P^{CP}_{\mu e}\right)
    = \Delta P^{CP}_{\mu e}(\delta)
      - \Delta P^{CP}_{\mu e}(\delta=0),
\end{eqnarray}
which isolates the dependence on the CP violating phase by subtracting the CP conserving reference value. To get over the problem of finding the source of CP violation (i.e. whether due to intrinsic CP phase and due to the matter effects), the observables $\delta\left(\Delta P^{CP}_{\mu e}\right)$ have been introduced which can prove useful to establish whether CP violation effects arise purely due intrinsic CP phase or a combination of intrinsic and extrinsic CP effects.
\begin{figure}[hbt!]
\centering
\vspace{-0.5cm}
\hspace{-1cm}
\includegraphics[width=0.594\textwidth]{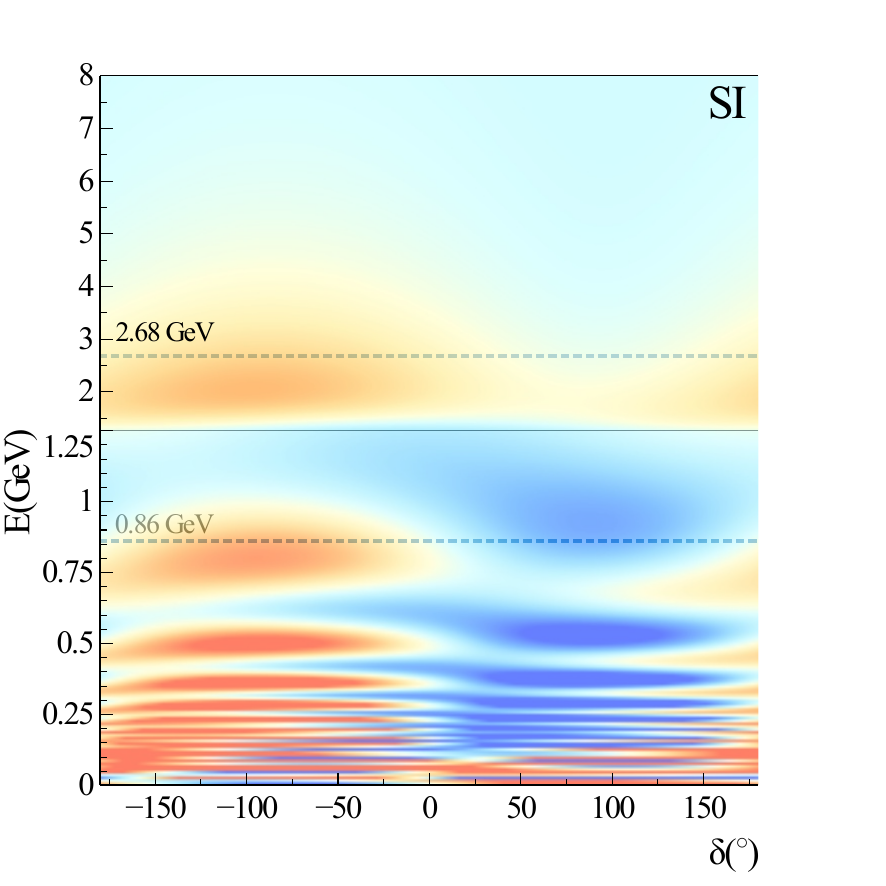}
\vspace{0.3cm}
\hspace{-2.55cm}
\includegraphics[width=0.593\textwidth]{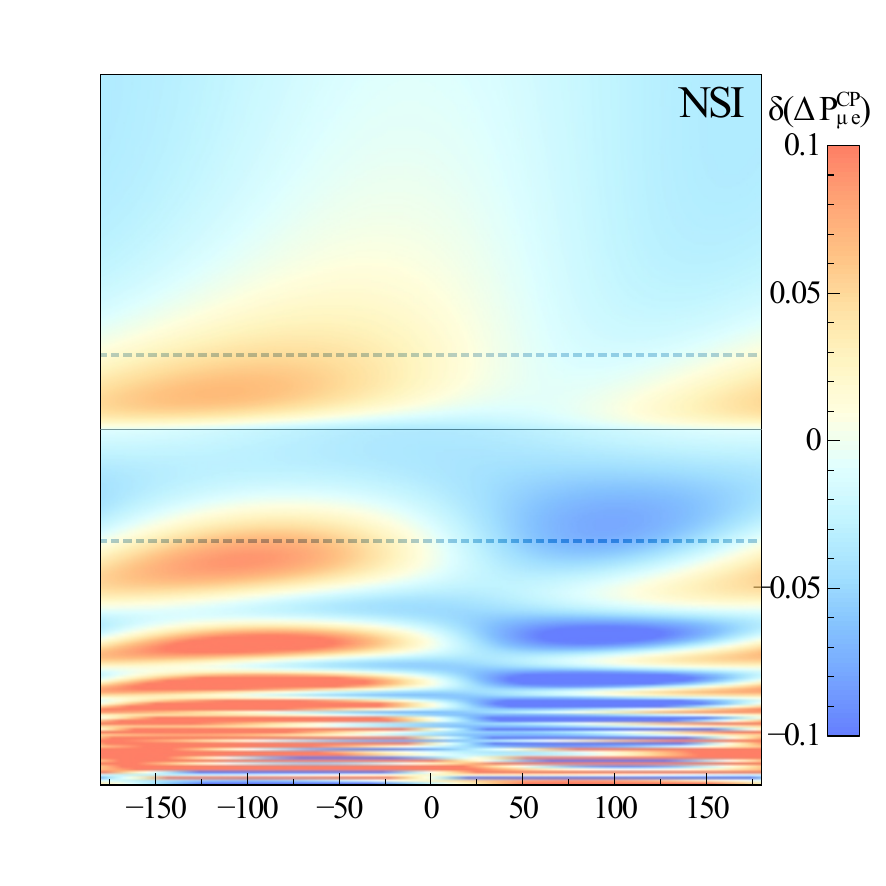}
\caption{\footnotesize{{{Oscillogram plot for $\delta\left(\Delta P^{CP}_{\mu e}\right)=\Delta P^{CP}_{\mu e}\left(\delta\right)-\Delta P^{CP}_{\mu e}\left(\delta=0\right)$ in E(GeV)-$\delta\left(^\circ\right)$ plane for standard matter effect (left panel) and  NSI  (right panel)}}}} 
\label{fig:6}
\end{figure}
This particular choice of observable $\delta\left(\Delta P^{CP}_{\mu e}\right)$ enables us to examine the possibility of extracting the information about the intrinsic CP phase. Subtracting the factor $\Delta P^{CP}\left(\delta=0\right)$ may eliminate the fake CP effect arising due to the purely matter effect. The remaining effects in the expression of CP violation are due to the intrinsic $\delta$. However, in the presence of NSI, due to the presence of the matter dependence term $\left(f^{2}+f^{\prime^{2}}\right)$, associated with $\cos{\delta}$, interference from fetching the absolute intrinsic $\delta$. This matter effect can be minimized if we probe the observable at the region where the matter effect is minimal, which can be achieved at the second oscillation maximum.

In Fig.~\ref{fig:4} we demonstrate the behaviour of this observable $\delta \left(\Delta P^{CP}_{\mu e}\right)$ as a function of the CP phase $\delta$ for energies corresponding to the first and second oscillation maxima, for DUNE.
The red curves correspond to the second oscillation maximum, while the blue curves represent the first. For each case, the dotted, dashed, and solid lines indicate the vacuum, standard matter, and NSI scenarios, respectively.
The pink shaded region identifies the parameter space, for $\delta=\pm 20^\circ$, where the extraction of the intrinsic CP phase remains feasible even in the presence of NSI. The red and blue shaded bands highlight the deviation from pure intrinsic CP violation induced by matter effects at the second and first oscillation maxima, respectively. The plot clearly shows that this deviation is minimized where the vacuum and NSI curves nearly overlap.

At the first oscillation maximum, around 2.68~GeV, the quantity $\Delta P^{CP}_{\mu e}$ exhibits a strong $\cos\delta$ type behaviour induced by matter effects, making it difficult to isolate the intrinsic CP phase. This matter induced contribution is substantially reduced at higher order maxima, where $\Delta P^{CP}_{\mu e}$ regains an approximately sinusoidal dependence on $\delta$. Consequently, the interference from the $\cos\delta$ terms is suppressed at the second oscillation maximum, allowing for a clearer and more robust determination of the CP phase.


The oscillogram plots in Fig.~\ref{fig:6} reinforce this conclusion even more clearly. Both the SI and NSI cases retain, to some extent, the characteristic features of the vacuum oscillogram. In particular, the approximate symmetry around the point of maximal CP violation at $\delta = 90^\circ$ is still visible, especially in the region of the second oscillation maximum.
Taken together, these observations highlight that, even in the presence of NSI, the second oscillation maximum offers a more reliable and experimentally accessible region for extracting the intrinsic CP phase. This makes it a particularly advantageous setting for precision studies of leptonic CP violation.
\label{theory}

{\section{ Experiment and Simulation details} 
\label{sec:3}
DUNE~\cite{DUNE:2015lol, DUNE:2021cuw} is designed as a next generation 
long baseline accelerator neutrino experiment, with neutrinos traveling approximately 
1300 km from Fermilab to the far detector located deep underground at the Sanford 
Underground Research Facility in South Dakota. Beyond the standard three-flavor oscillation scenario, DUNE’s long baseline also makes it highly sensitive to possible NSI. As per the discussion in Section~\ref{sec:2}, NSI can generate parameter degeneracies such as mixing the standard CP phase $\delta$ with NSI phases and thereby shift or reduce DUNE’s apparent sensitivity to CP violation~\cite{Masud:2016bvp}. Also, the second oscillation maximum may play a crucial role in getting the CP violation sensitivity and disentangling the intrinsic contribution from the extrinsic for both SI and NSI scenarios. Therefore, we use the GLoBES framework~\cite{Huber:2004ka, Huber:2007ji} together with the NSI analysis tool~\cite{kopp2010sterile} for performing the simulations. We assume a 40 kton liquid argon time projection chamber (LArTPC) as the far detector. The TDR beam design employs 120~GeV protons at 
1.2~MW beam power, corresponding to an annual exposure of $1.1\times10^{21}$ protons 
on target (POT), which explores the first oscillation maximum, whereas the 8~GeV beam explores the second oscillation maximum with 
3~MW power, yielding $40.1\times10^{21}$ POT per year. The 120~GeV beam seems to be optimized around 2.5 GeV, i.e., near the first oscillation maximum, and the 8 GeV beam seems to be optimized around the second oscillation maximum. The muon-neutrino flux in neutrino (solid line) and antineutrino (dotted line) modes is shown in Fig.~\ref{dune_flux}, where the red and blue curves denote the 120 GeV and 8 GeV beams, respectively. 
We adopt a balanced runtime of 
3.5 years each in neutrino and antineutrino modes, leading to exposures of about 336 and 
840~kton-MW-years for the 120~GeV  and 8~GeV  beams, respectively. Notably, the staged 
seven-year TDR run also achieves a total exposure of 336~Kton-MW-years~\cite{DUNE:2020jqi}. The 
details of systematic uncertainties and detection efficiencies for the TDR 
configurations are available in Refs.~\cite{DUNE:2021cuw}.

\vskip 0.5cm
\noindent 
For all our numerical simulations, we employ a constant Earth matter density of $\rho = 2.484\,\mathrm{g/cm^3}$, following the PREM profile~\cite{Stacey1977, Dziewonski1981}. Unless otherwise specified, we use the benchmark values for both standard and non-standard neutrino oscillation parameters from Table~\ref{oscparams}. Throughout this work, we adopt the normal mass ordering (NO) as our default assumption. We marginalize over all standard oscillation parameters (except the Dirac CP violating phase $\delta$ and the NSI parameters) to determine the sensitivity. Following experimental specifications, we include a 2\% systematic uncertainty on the average matter density, as outlined in the Technical Design Report (TDR).
    \begin{figure}[t!]
   \centering
\includegraphics[width=4.2in]{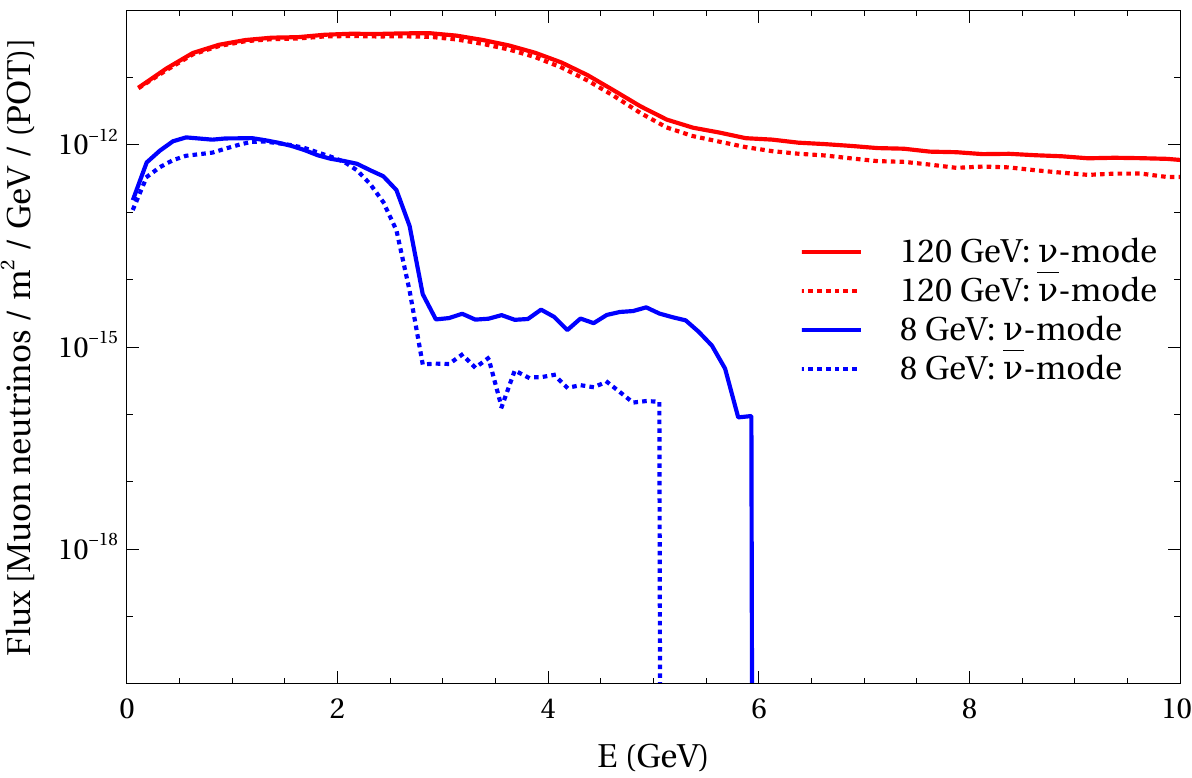}
\caption {The red and blue curves represent the muon neutrino flux for 120 GeV beam~\cite{DUNE:2020ypp, DUNE:2021cuw} and 8 GeV beam, respectively. The solid and
dotted curves indicate flux in neutrino and antineutrino modes, respectively.}
\label{dune_flux}
\end{figure} 
\section{Statistical Analysis}
\label{sec:4}
\subsection*{Event Spectrum at DUNE}
In this section, we present the expected event spectra at the DUNE far detector for both the SI and NSI scenarios using the GLoBES software.
The spectra are shown for a 3.5 year runtime, each for FHC (forward horn current), i.e., neutrino mode, and RHC (reverse horn current), i.e., antineutrino mode, for a total runtime of seven years for both 120 GeV and 8 GeV beams. One of the event spectrum plots is also shown with the
optimized runtime combination of both the above mentioned beams, keeping the total runtime fixed, which is discussed in detail in Section~\ref{sec:5}.
 Fig. \ref{event_120} and Fig. \ref{event_8} ~show the expected rate of selected events for $\nu_{e}$ (top panel) and $\bar{\nu}_{e}$ (bottom panel) appearance for  of SI (left panel) and NSI (right panel) scenarios, considering expected flux, cross section, and oscillation probabilities as a function of reconstructed neutrino energy at a baseline of 1258 km. 
In general, one notes that the event spectra are peaked at the value of energy where the flux is optimized. For the 120 GeV beam at the first oscillation maximum, the events peak around $2.6$ GeV, while for the 8 GeV beam at the second oscillation maximum, it is around $0.8-0.9$ GeV. The signal for $\nu_e$ appearance is an excess of charged current (CC) $\nu_e$ and $\bar\nu_e$ interactions over the
expected background in the far detector. The background to the $\nu_e$ appearance is composed of
\begin{itemize}
\item Beam  $(\nu_e + \bar\nu_e)$ CC : CC interactions of $\nu_e$ and $\bar\nu_e$ intrinsic to the beam; 
\item 
$ (\nu_\tau + \bar\nu_\tau)$ CC :    $\nu_\tau$ and $\bar\nu_\tau$ CC events in which the $\tau$'s decay leptonically into electrons/positrons.
  \item  $(\nu_\mu + \bar \nu_\mu) $ CC :   misidentified $\nu_\mu$ and $\bar\nu_\mu$ CC events; 
\item NC: neutral current backgrounds.
\end{itemize} 

\begin{figure}[h!]
\centering
\includegraphics[width=0.9\textwidth]{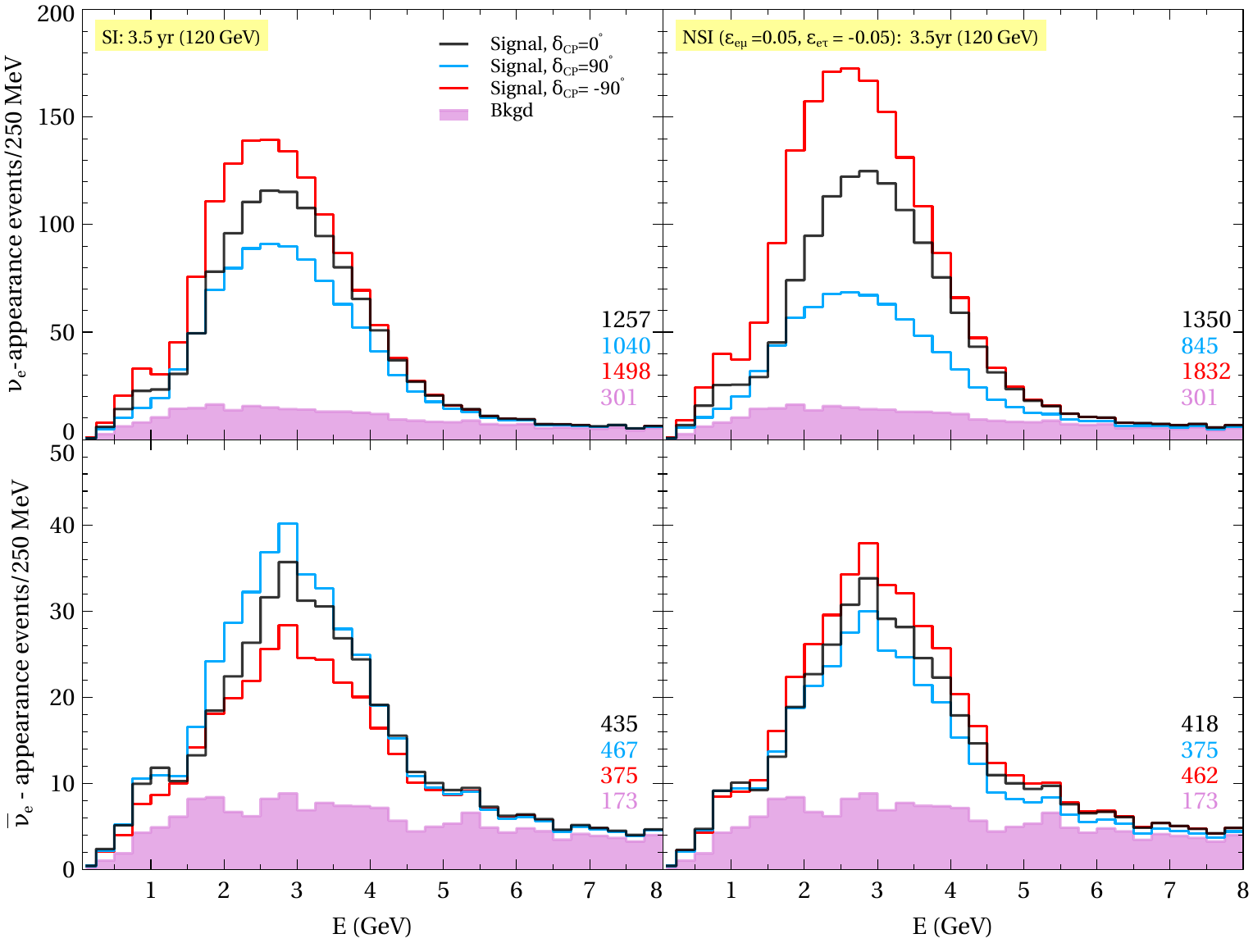}
\caption{\footnotesize{{{Event spectra of 120 GeV beam for $\nu_{e}$ appearance (top panel)  and $\bar{\nu}_{e}$ appearance (bottom panel) with SI (left panel) and NSI (right panel) with a runtime of 3.5 years in each mode.}}
  }} 
\label{event_120}
\end{figure} 
 \begin{figure}[h!]
\centering
\includegraphics[width=0.9\textwidth]{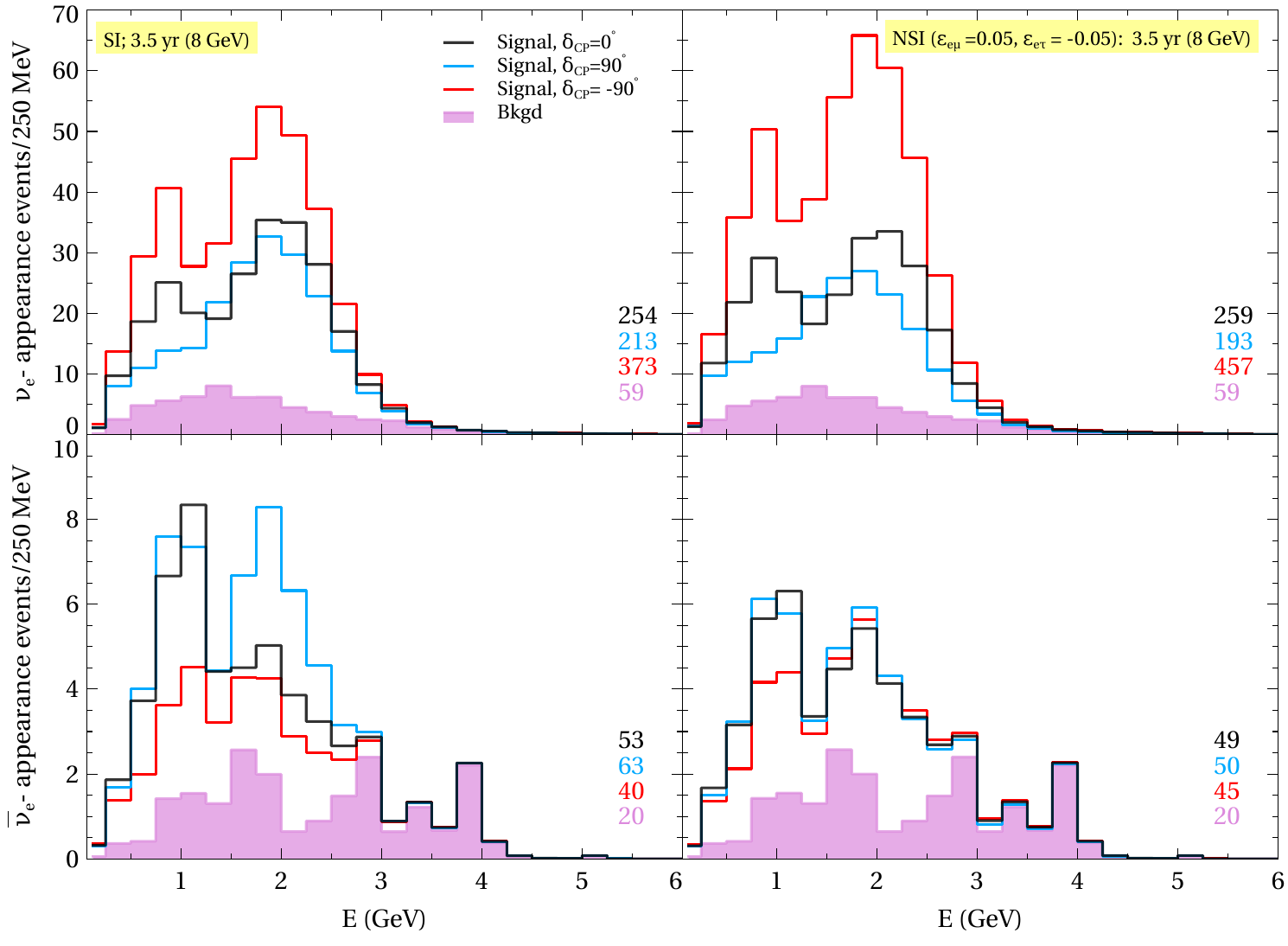}
\caption{\footnotesize{{{Event spectra of 8 GeV beam for $\nu_{e}$ appearance (top panel)  and $\bar{\nu}_{e}$ appearance (bottom panel) with SI (left panel) and NSI (right panel) with a runtime of 3.5 years in each mode.}}
  }} 
\label{event_8}
\end{figure} 
  \begin{figure}[h!]
\centering
\includegraphics[width=0.9\textwidth]{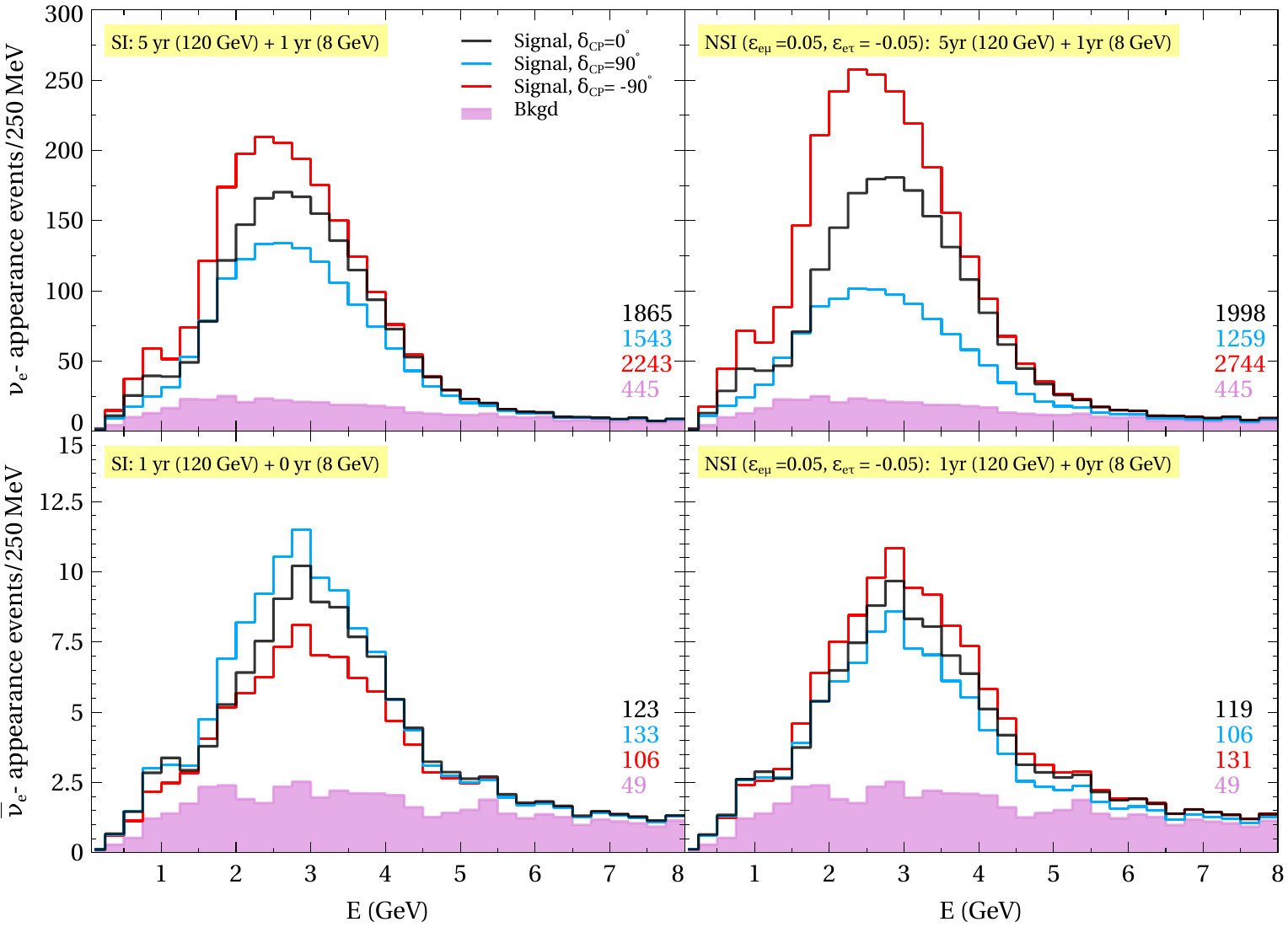}
\caption{\footnotesize{{{Event spectra of combination of 8 GeV and 120 GeV beams for $\nu_{e}$ appearance (top panel)  and $\bar{\nu}_{e}$ appearance (bottom panel) with SI (left panel) and NSI (right panel) with runtimes as mentioned in the figure.}}
  }} 
\label{event_8+120}
\end{figure} 
  
\begin{table}[h]
\scriptsize
{ \centering
 \hspace{0cm}
	\begin{tabular}{|l | c | c | c  |}
\hline
 {\bf{Physics Scenarios}}   & \multicolumn{1}{c |}{120 GeV:  $\nu_{e}$ ($\bar\nu_{e}$) app }  & \multicolumn{1}{c |}{8 GeV:  $\nu_{e}$ ($\bar\nu_{e}$) app } & \multicolumn{1}{c |}{120 GeV + 8 GeV:  $\nu_{e}$ ($\bar\nu_{e}$) app } \\

\hline
{\colorbox{pink}{\bf{Event rate for SI}}} &&&\\
Signal $\delta= 0$                  & 1257 (435) &  254 (53) & 1865 (123) \\
Signal $\delta= \pi/2$              & 1040 (467) &  213 (63) & 1543 (133) \\
Signal $\delta= -\pi/2$             & 1498 (375) &  373 (40) & 2243 (106) \\
Total Bkgd                          & 301 (173)   &  59 (20) & 445 (49) \\
\hline
{\colorbox{pink}{\bf{Event rate for NSI}}} &&&\\
Signal $\delta= 0$                  & 1350 (418) &  259 (49) & 1998 (119) \\
Signal $\delta= \pi/2$              & 845 (375) &  193 (50) & 1259 (106) \\
Signal $\delta= -\pi/2$             & 1832 (462) &  457 (45) & 2744 (131)\\
Total Bkgd                          & 301 (173)  & 59 (20) & 445 (49) \\
\hline
\end{tabular}
	\caption{\footnotesize{Total number of event rates of  appearance channel for all three beam options with SI and NSI.}}
	\label{ev_tab}

   }
\end{table}
\noindent

In each event spectrum figure, we display the signal for $\delta=0^\circ$, $\pi/2$, and $-\pi/2$, along with the total backgrounds. We can clearly see that the event rate for $\delta=-\pi/2$ is larger than for $\delta=\pi/2$ in neutrino mode, with the reverse being true for antineutrino mode. Both the signal and the backgrounds are larger in neutrino mode than in antineutrino mode.
In the presence of NSI ($\epsilon_{e\mu} = 0.05$ and $\epsilon_{e\tau}=-0.05$), the signal increases for $\delta=-\pi/2$ and decreases for $\delta=\pi/2$, regardless of the beam mode. Statistically, the 8 GeV beam performs worse than the 120 GeV beam. However, due to its importance in determining CP violation sensitivities at the second oscillation maximum, combining the two beams may yield better results than using the standard 120 GeV beam alone.
The optimal runtime combination for both beams, determined by optimizing the CP violation sensitivity at DUNE (see Section~\ref{sec:5}), is found to be 5 years in neutrino mode and 1 year in antineutrino mode for the 120 GeV beam, plus 1 year in neutrino mode only for the 8 GeV beam. This keeps the total runtime fixed at 7 years. The event rate for the combined beams is shown in Fig.~\ref{event_8+120}.

\noindent
Table \ref{ev_tab} shows the integrated event rates for the total signal and background in the $\nu_{e}$ ($\bar\nu_{e}$) appearance channel in neutrino (antineutrino) mode, considering both SI and NSI. The total rates are integrated over the reconstructed neutrino energy range used in this analysis, 0.5-8 GeV.
The event spectra for SI and NSI clearly show that NSI significantly increases the $\nu_{e}$ appearance event rate. However, the rate of increment is smaller in the $\bar{\nu}_{e}$ appearance channel, implying that NSI has a more pronounced effect on neutrinos than on antineutrinos compared to the SI case. For the NSI case at $\delta = -\pi/2$, the signal improves relative to SI by 334 (87), 84 (5), and 501 (25) events for the 120 GeV, 8 GeV, and combined beams, respectively, in neutrino mode (antineutrino mode). In contrast, a reduction occurs at $\delta_{\text{CP}} = \pi/2$, with 195 (92), 20 (13), and 284 (27) fewer events for the same beam configurations.

\section{Sensitivity Analysis}
\label{sec:5}
DUNE is designed to explore leptonic CP violation within the standard three flavor neutrino oscillation framework. 
A conclusive signal of CP violation requires the true value of $\delta$ to differ significantly from the CP conserving values $0$ and $\pi$. To quantify the experiment’s ability to discriminate between CP violating and CP conserving scenarios, we define the statistical measure $\chi^{2}$ as  
\begin{equation}
    \chi^{2} \equiv \min_{\delta_{\mathrm{test}}}
    \sum_{i=1}^{x}\sum_{j=\nu,\bar{\nu}}
    \frac{\big[N^{i,j}_{\mathrm{true}}(\delta_{\mathrm{true}})
    -N^{i,j}_{\mathrm{true}}(\delta_{\mathrm{true}}=0,\pi)\big]^{2}}
    {N^{i,j}_{\mathrm{true}}(\delta_{\mathrm{true}})}.
    \label{delchi}
\end{equation}

\noindent
In this expression, $N^{i,j}_{\mathrm{true}}$ and $N^{i,j}_{\mathrm{test}}$ represent the predicted event counts for the $(i,j)$th energy bin and neutrino/antineutrino mode, respectively. The index $i$ runs over the reconstructed energy bins ($i=1,\ldots, N_x$), where the binning scheme is experiment dependent. For DUNE, the binning consists of 64 bins of width 125 MeV in the 0-8 GeV region, along with seven wider bins covering 8-20 GeV. The second index $j$ sums over both neutrino and antineutrino channels~\cite{Rout:2020emr}.
It is important to note that we employ the Pearson definition of the test statistic, for which $\Delta\chi^{2} = \chi^{2}$ and the corresponding significance is given by $N_{\sigma} = \sqrt{\Delta\chi^{2}}$. This choice is appropriate since no statistical fluctuations are included in the simulated event samples. For sufficiently large data sets, the alternative definition based on the log-likelihood formalism yields numerically comparable results.

\medskip

The CP violation discovery potential for a given true value of the phase, $\delta_{\mathrm{true}}$, is evaluated through the quantity  
\begin{eqnarray*}
    \Delta\chi^{2}
   & = \min\!\left[\Delta\chi^{2}(\delta_{\mathrm{true}}=0),
                 \,\Delta\chi^{2}(\delta_{\mathrm{true}}=\pi)\right]~, ~\mathrm{where}\quad\Delta\chi^{2} & =\chi^{2}_{\delta_{\mathrm{test}}}
               -\chi^{2}_{\delta_{\mathrm{true}}}.
\end{eqnarray*}
A larger value of $\Delta\chi^{2}$ indicates a greater ability to reject the CP conserving hypotheses and therefore a stronger sensitivity to leptonic CP violation.

To evaluate the sensitivity to CP violation, we construct the test statistic $\chi^{2}$ under the assumption of CP conservation. In this framework, the test values of the CP violating phase are fixed to $\delta = 0$ or $\pi$, representing the two CP conserving hypotheses. The corresponding $\chi^{2}$ is then computed for each possible true value of the phase within the full physical interval $\delta \in [-\pi,\pi]$. This approach enables us to quantify how effectively a given experimental setup can distinguish CP violating values of $\delta$ from the CP conserving cases.

The characteristic double peak structure visible in Fig.~\ref{cp_sens} reflects the minimization over the test parameters for a specific set of true oscillation parameters. This procedure is repeated over the entire range of true parameter values.
In addition, the total sensitivity is obtained by summing the $\chi^{2}$ contributions from both the appearance channel $(\nu_\mu \rightarrow \nu_e)$ and the disappearance channel $(\nu_\mu \rightarrow \nu_\mu)$. Including both channels ensures that the sensitivity estimate fully incorporates the complementary information provided by the two oscillation modes.
\subsection{CP violation sensitivity}
With the event rates and spectral features characterized, we turn to a key objective of long-baseline experiments: probing CP violation in the neutrino sector. We analyze DUNE’s capability to measure the CP violating phase $\delta$ and study how the presence of NSI parameters modifies this sensitivity. The left panel of Fig.~\ref{cp_sens} shows the resulting CP violation sensitivity as a function of the true value of $\delta$ for several representative beam configurations. The blue curve corresponds to the 8~GeV beam, while the red and green curves correspond to the 120~GeV and combined (8 + 120)~GeV beam options, respectively. 
Throughout this analysis, we adopt $\varepsilon_{e\mu}=0.05$ and $\varepsilon_{e\tau}=-0.05$ as benchmark NSI values and marginalize over the standard oscillation parameters when computing the sensitivity.
As expected, the sensitivity peaks near $\delta = \pm 90^\circ$, the points of maximal CP violation, for both SI (dashed curves) and NSI (solid curves). This reflects the larger statistical separation between the event spectra at $\delta = \pm 90^\circ$ and those at the CP conserving values $\delta = 0$ or $\pi$.
\begin{figure}[hbt!]
\centering
\hspace*{-0.95cm}
\includegraphics[width=0.50\textwidth]{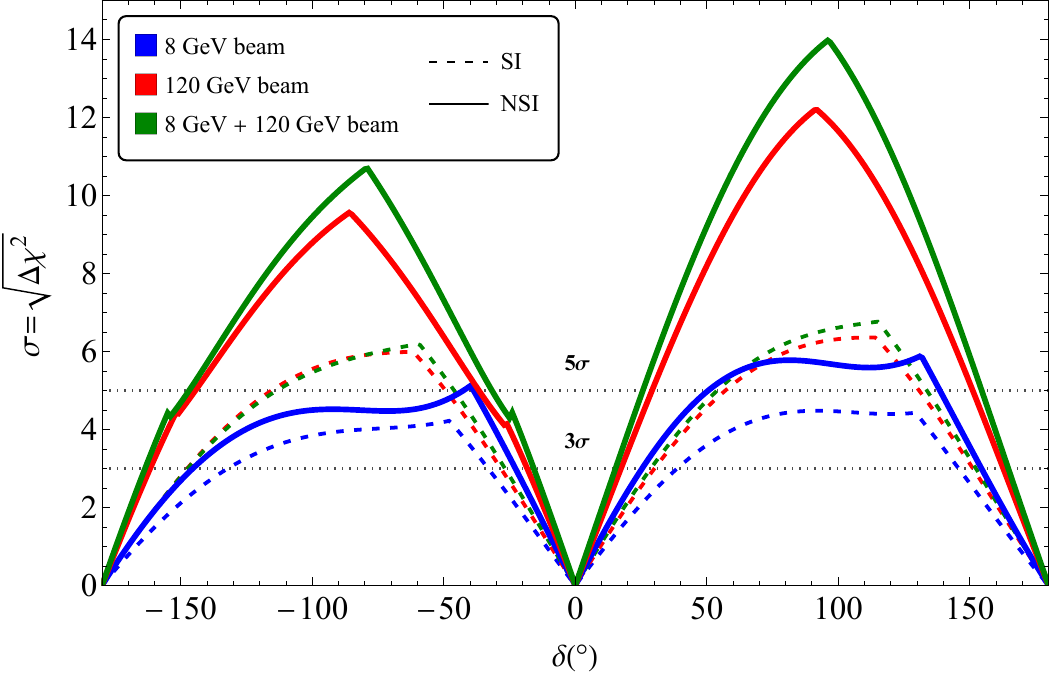}
\hspace*{-0.1cm}
\includegraphics[width=0.51\textwidth]{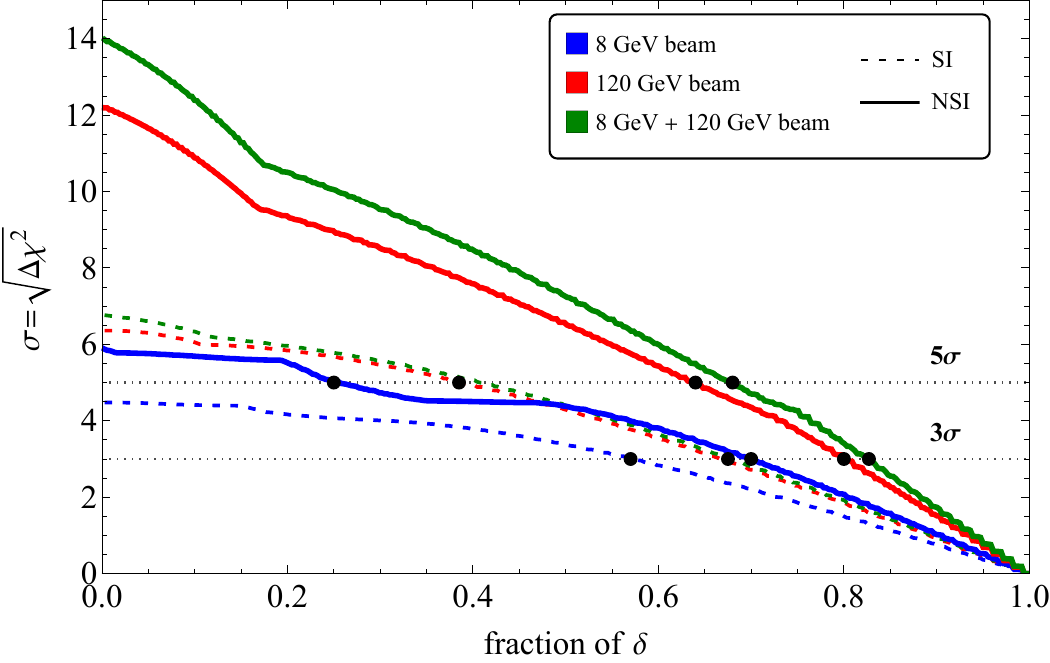}
\caption{\footnotesize{{{ Significance of DUNE determination of CP sensitivity (left panel) and fractional CP sensitivity (right panel) as a function of the true value of $\delta$, assuming seven years of exposure equally distributed between neutrino and antineutrino modes. The dashed curves correspond to the standard interaction (SI) scenario, while the solid curves represent the case including non-standard interactions (NSI). Results are shown for three different beam configurations: $8~\mathrm{GeV}$ (blue), $120~\mathrm{GeV}$ (red), and the combined $(8+120)~\mathrm{GeV}$ setup (green).}}
  }} 
\label{cp_sens}
\end{figure}
The 8~GeV beam, which mainly probes the second oscillation maximum, does not reach the $5\sigma$ level across the full range of $\delta$ under SI, even with improved energy resolution. However, in the presence of NSI, the same beam achieves sensitivities slightly above $5\sigma$, underscoring the enhanced CP violating effects induced by NSI at lower energies. The 120 GeV beam, centered near the first oscillation maximum, provides a significantly stronger CP violation reach, exceeding $5\sigma$ for SI and surpassing $10\sigma$ for the benchmark NSI scenario.

A notable improvement occurs when the 8~GeV and 120~GeV beams are combined (green curve). Using 6 years of exposure for the 120~GeV beam (5 years in $\nu$ mode and 1 year in $\bar{\nu}$ mode) together with 1 year of 8~GeV running (1 year in $\nu$ mode), the CP violation sensitivity experiences a substantial enhancement in the presence of NSI. While this configuration does not yield a significant improvement beyond the 120~GeV beam alone for the SI case, it becomes essential once NSI effects are included, pushing the sensitivity beyond the $12\sigma$ level. This reflects the complementary roles played by the first and second oscillation maxima, demonstrating the advantage of a dual-beam strategy for maximizing CP violation discovery potential in NSI scenarios. 
\subsection{Fractional CP violation sensitivity}
The right panel of Fig.~\ref{cp_sens} presents the same sensitivity information as the left panel, but expressed in terms of the fraction of CP phase values for which CP violation can be established at a given confidence level. This representation provides a complementary perspective by quantifying how much of the $\delta$ parameter space is experimentally accessible.

For the 8~GeV beam (blue curve), the SI scenario (dashed line) allows CP violation to be discovered at the $3\sigma$ level for about $56\%$ of all possible $\delta$ values, though no region reaches $5\sigma$. In contrast, when NSI are included (solid line), the same beam configuration achieves $5\sigma$ discovery for roughly $24\%$ of the $\delta$ range, reflecting the enhanced CP violating effects induced by NSI near the second oscillation maximum.

For the SI case, the 120~GeV beam (red curve) and the optimized combined 8~GeV + 120~GeV configuration (green curve) yield similar performance. CP violation can be established at the $3\sigma$ ($5\sigma$) level for about $67\%$ ($40\%$) of the $\delta$ values. However, the impact of NSI is much more pronounced. With NSI, the 120~GeV beam alone provides sensitivity at the $3\sigma$ ($5\sigma$) level over approximately $80\%$ ($64\%$) of the parameter space, while the combined configuration extends this coverage to nearly $85\%$ ($70\%$).

These results align with the event rate analysis. The presence of NSI enhances the $\nu_e$ appearance rate in the neutrino mode by modifying the matter affected oscillation probabilities. This enhancement directly strengthens the CP violation discovery reach, particularly for the combined (8 + 120)~GeV beam, where the contributions from the first and second oscillation maxima complement each other most effectively. Overall, the fractional coverage perspective underscores the robustness of DUNE’s CP violation sensitivity and highlights the importance of optimized dual beam strategies, especially in scenarios where NSI effects are present.
\section{Conclusion}
\label{sec:6}
The present study investigates the impact of NSI in the context of DUNE, specifically focusing on the extraction of intrinsic CP violation effects within the three flavor neutrino oscillation framework. A key aspect of this analysis is the significance of the second oscillation maximum in the measurement of the CP violating phase $\delta$ in the presence of NSI. Our results demonstrate that, in the presence of NSI, the oscillation probability $P_{\mu e}$ for the channel $\nu_{\mu}\rightarrow\nu_{e}$ at the second oscillation maximum exhibits a significant enhancement compared to both the vacuum and standard matter interaction cases. This enhancement directly correlates with an increase in the CP violation term $\Delta P^{CP}_{\mu e}$,  as illustrated in Fig. \ref{fig:oscideldelp}. The figure explicitly highlights that the inclusion of NSI amplifies CP violation most prominently in the region of the second oscillation maximum. Consequently, this region emerges as an optimal probe for investigating CP violation in the neutrino sector under NSI effects. 
However, it is essential to recognize that CP violation in neutrino oscillations is inherently entangled with matter effects. This interplay presents a challenge in isolating the intrinsic CP violation component, particularly in the presence of NSI. Previous studies have explored the extraction of intrinsic CP violation under standard matter interactions, leveraging the parameter, ${\delta(\Delta P_{\mu e})=\Delta P_{\mu e}(\delta=\pi/2)-\Delta P_{\mu e}(\delta=0)}$ to eliminate fake CP effects induced by matter interactions. While this approach enhances the prospects for identifying intrinsic CP violation, the incorporation of NSI complicates the extraction process, as clearly depicted in the oscillogram plots of Fig. \ref{fig:3}. Despite the disruption of the symmetric structure of intrinsic CP violation under NSI, our analysis suggests that its extraction remains viable, particularly in the vicinity of CP conserving regions. To facilitate this, we introduce the parameter $\delta(\Delta P^{CP})=\Delta P^{CP}(\delta)-\Delta P^{CP}(\delta=0)$, which enhances the feasibility of isolating intrinsic CP violation. Fig. \ref{fig:4} further reinforces the advantage of the second oscillation maximum in minimizing matter effects. Notably, in the neighborhood of the CP conserving region $(\delta=\pm 20^\circ)$, our results indicate a promising avenue for extracting intrinsic CP violation even in the presence of NSI.
\\ The probability level analysis reveals that, in the presence of NSI, the appearance probability of electron neutrinos is significantly enhanced at $\delta=-90^\circ$. This enhancement has a pronounced impact on the CP violating term $\Delta P_{\mu e}$,  which attains its maximal value at $\delta=-90^\circ$. The effect progressively amplifies from vacuum to standard matter interactions and further in the presence of NSI.
However, at $\delta=-90^\circ$, the extraction of intrinsic CP violation becomes increasingly challenging due to the entanglement of matter effects with the CP phase. This complexity arises from the difficulty in isolating the genuine CP violating contribution from the fake CP effects induced by the matter potential. Consequently, while maximal CP violation can be achieved at $\delta=-90^\circ$ in the region of the second oscillation maximum under NSI, precise determination of  $\delta$ remains nontrivial.
Nevertheless, the vicinity of the CP conserving region $(\delta=\pm 20^\circ)$ provides a viable avenue for extracting information about the intrinsic CP phase. In this region, the matter effects are relatively minimized, thereby improving the prospects for disentangling the intrinsic CP violation from background contributions.
\\ The statistical analysis corroborates this observation. In the presence of SI, approximately 1478 $\nu_{e}$ events are observed at $\delta=-90^\circ$. This number is effectively reduced to 496 in the $\nu$ mode in the presence of NSI, indicating a significant effect.
The effective reduction in the event rate strengthens the prospects of probing CP violation under the influence of NSI, offering a more comprehensive understanding of the underlying physics.
\\ 
We find that DUNE can achieve high significance sensitivity to leptonic CP violation over a wide range of the Dirac phase. In the SI scenario (dashed line), the $120$~GeV beam (red line in Fig. \ref{cp_sens}) alone surpasses $5\sigma$ sensitivity near $\delta=\pm\pi/2$, while the $8$~GeV beam (blue line in Fig. \ref{cp_sens}) contributes mainly at the second oscillation maximum without reaching discovery level. In the presence of NSI (solid line), however, the $120$~GeV beam yields sensitivity above $10\sigma$, and the optimized $(8+120)$~GeV configuration (green line in Fig. \ref{cp_sens}) further extends the coverage beyond $12\sigma$, enabling CP violation discovery for nearly $70\%$ of the $\delta$ parameter space at $5\sigma$. These results establish the complementarity of the first and second oscillation maxima and underscore the importance of optimized combined beam strategies for maximizing DUNE’s discovery potential, especially under nonstandard interaction effects.
\\
The fractional sensitivity study shows that DUNE can establish leptonic CP violation over a substantial portion of the parameter space. In the SI case, the $120$~GeV beam alone covers about $40\%$ of $\delta$ values at $5\sigma$, while in the presence of NSI, this coverage increases to nearly $64\%$. The optimized $(8+120)$~GeV beam configuration further enhances discovery potential, extending the $5\sigma$ reach to about $70\%$ of the $\delta$ range. These results demonstrate the robustness of DUNE’s capability to probe CP violation and highlight the essential role of combined beam strategies, particularly when NSI effects are considered.

In summary, our study demonstrates that while NSI substantially modify the extraction of intrinsic CP violation, the second oscillation maximum at DUNE remains a powerful probe to recover sensitivity. These results underscore the necessity of carefully incorporating NSI effects in future precision analyses and highlight the complementary role of multiple oscillation maxima in advancing our understanding of CP violation in the neutrino sector.
\section*{ACKNOWLEDGMENTS}
RB would like to thank the Ministry of
Electronics and IT for the financial support through the Visvesvaraya fellowship scheme for carrying out this research work. The use of HPC cluster at SPS, JNU funded by DST-FIST is acknowledged. JR would like to thank OSHEC
(Odisha State Higher Education Council) for the financial support through Mukhyamantri Research Innovation (MRI) for Extramural Research Funding 2023.  SP acknowledges the financial support under MTR/2023/000687 funded by SERB, Government of India. 
The research of PM is supported in part by the Inter-University Centre for Astronomy and Astrophysics (IUCAA), Pune through its
Associateship Programme. PM would like to acknowledge IIT Kanpur for kind hospitality during the finishing stages of this work.
\\
This work reflects the views of the authors and not those of the DUNE collaboration.
\appendix
\section*{Appendix}
\section{Standard three flavor neutrino oscillation}
\label{stdosc}
In the standard three flavor framework, the evolution equation for neutrinos is
 \begin{eqnarray}
   i\dfrac{d}{dt} \begin{pmatrix}
\nu_e  \\
\nu_{\mu}  \\
\nu_{\tau} 
\end{pmatrix} &=& \left[\frac{1}{2E} U \begin{pmatrix}
0 & 0 & 0  \\
0 & \Delta m_{21}^2 & 0 \\
0 & 0 & \Delta m_{31}^2
\end{pmatrix} U^{\dagger} + \begin{pmatrix}
V_{CC} & 0 & 0  \\
0 & 0 & 0 \\
0 & 0 & 0
\end{pmatrix}\right]\begin{pmatrix}
\nu_e  \\
\nu_{\mu}  \\
\nu_{\tau} 
\end{pmatrix}~.
\label{eq:1}
\end{eqnarray}
Here, $\Delta m_{ij}^2 = m_i^2-m_j^2$ denotes the mass-squared difference  with  $i,j=1,2,3$ ($i \neq j$), $E$ denotes the neutrino energy, and $V_{CC}$  is the charged current (CC) matter potential defined by
\begin{equation}
V_{CC} = \sqrt{2} G_F N_e \simeq 7.6 \times Y_e \times \frac{\rho}{10^{14}
\mbox{g}/\mbox{cm}^3} \, \mbox{eV}\,,
\label{eq:3}
\end{equation}
where $G_F$ is the Fermi coupling constant,  $\rho$ is the matter density, and 
$Y_e = {N_e}/{(N_p + N_n)}$ is the 
electron fraction. $N_e$ is the electron number density while $N_p$ and $N_n$ are the proton and neutron number densities respectively. For an electrically neutral and isoscalar medium, 
$N_e = N_p = N_n$ and then $Y_e \simeq 0.5$ \cite{Chatterjee:2014gxa}. Earth matter density, which for DUNE we take as $\rho=$ 2.848 $\mbox{gm}/\mbox{cm}^{3}$.
\section{Brief review of $\Delta P^{CP}_{\mu e}$ in vacuum and matter}
\subsubsection*{$\Delta P^{CP}_{\mu e}$ in vacuum }
\label{CPvac}
In absence of the matter potential the factor $f$ takes the form simply $f_{v}=f\equiv \sin{\Delta}$ modifying the $P_{0}$ term to $P_{0}^{v}=4s^{2}_{13}s^{2}_{23}f^{2}_{v}$. 
The probability in vacuum can thus be written as
\begin{equation*}
	P_{\mu e}= 4 s_{13}^2 s^2_{23}  \sin^2{\Delta}
	 + 8 s_{13} s_{12} c_{12} s_{23} c_{23} \alpha \sin{\Delta} \cos({\Delta + \delta})
\label{eq:16}
\end{equation*}
where the factor $g\rightarrow 1$ in the vacuum limit, and $f=\frac{\sin[(1-v)\Delta]}{1-v}\rightarrow \sin{\Delta}$. Similarly, the appearance probability for antineutrinos, accounting for $\delta$, flip $\delta\rightarrow -\delta$ is given by 
\begin{equation*}
	\bar{P}_{\mu e}=4 s_{13}^2 s^2_{23}  \sin^2{\Delta}
	+ 8 s_{13} s_{12} c_{12} s_{23} c_{23} \alpha \sin{\Delta} \cos({\Delta - \delta})
\label{eq:17}
\end{equation*}
Therefore, the intrinsic CP violation in neutrino oscillation in vacuum is given by
\begin{equation}
    \Delta P^{CP}_{\mu e}=16 s_{13} s_{12} c_{12} s_{23} c_{23} \alpha \sin^{2}{\Delta}\sin{\delta}
\end{equation}
 It is evident that the expression for CP violation in vacuum, representing the intrinsic CP violation in the neutrino sector, exhibits a sinusoidal dependence on the CP phase $\delta$ \cite{Rout:2020emr, Gago:2009ij}.
 \subsubsection*{$\Delta P^{CP}_{\mu e}$ in  matter}
 \label{CPmat}
With the inclusion of matter effects, the factor $f$ in the transition probability expression for neutrinos modifies as  $f=\sin{[(1-v)\Delta]}/(1-v)$ and for the antineutrino $f^{\prime}=\sin{[(1+v)\Delta]}/(1+v)$. 
Meanwhile, the factor $g=\sin{v\Delta}/v$ remains unchanged for both neutrinos and anti-neutrinos. Therefore, in the presence of matter potential, the neutrino appearance probability takes the form 
\begin{equation*}
 	P_{\mu e} = P_{0} + P_{1} = 4 s_{13}^2 s^2_{23}  f^2 + 8 s_{13} s_{12} c_{12} s_{23} c_{23} \alpha f g \cos({\Delta + \delta}) 
 \label{eq:19}
 \end{equation*}
while the anti-neutrino probability is given by
 \begin{equation*}
 	\bar{P}_{\mu e} =  4 s_{13}^2 s^2_{23}  (f^{\prime})^2 + 8 s_{13} s_{12} c_{12} s_{23} c_{23} \alpha f^{\prime} g \cos({\Delta - \delta}). 
 \label{eq:20}
 \end{equation*}
 The CP asymmetry in the presence of matter effects is thus expressed as
 \begin{equation}
     \Delta P^{CP}_{\mu e}=4 s_{13}^2 s^2_{23}(f^{2}-f^{\prime^{2}})+ 8 s_{13} s_{12} c_{12} s_{23} c_{23} g\bigg[\cos{\Delta}(f-f^{\prime})\cos{\delta}-(f+f^{\prime})\sin{\Delta}\sin{\delta}\bigg]
 \end{equation}
 In the presence of SI with the matter potential, the CP violation term exhibits both $sine$ and $cosine$ dependencies on the CP phase $\delta$.  However, at the oscillation maxima, the $cosine$ contribution vanishes, resulting in $\Delta P^{CP}_{\mu e}$ showing a pure dependence $sinusoidal$ on the intrinsic CP phase $\delta$.
\bibliographystyle{jhep}
\bibliography{nsi}

\end{document}